\documentclass[%
 reprint,
superscriptaddress,
 amsmath,amssymb,
 prl, nolongbibliography,
]{revtex4-2}

\usepackage[english]{babel} 

\usepackage{graphicx}% Include figure files
\usepackage{dcolumn}% Align table columns on decimal point
\usepackage{bm}% bold math
\usepackage{hyperref}% add hypertext capabilities
\usepackage{xcolor}
\hypersetup{
    colorlinks,
    linkcolor={red!50!black},
    citecolor={blue!50!black},
    urlcolor={blue!80!black}
}

\begin{document}

\title{Geometry-sensitive protrusion growth directs confined cell migration}% Force line breaks with \\

\author{Johannes Flommersfeld}
\affiliation{Department of Physics and Astronomy, Vrije Universiteit Amsterdam, 1081HV Amsterdam, Netherlands}%
\affiliation{Arnold Sommerfeld Center for Theoretical Physics and Center for NanoScience, Department of Physics, Ludwig-Maximilian-University Munich, Theresienstraße 37, D-80333 Munich, Germany}
\author{Stefan Stöberl}
\affiliation{Faculty of Physics and Center for NanoScience, Ludwig-Maximilian-University, Geschwister-Scholl-Platz 1, D-80539 Munich, Germany}
\author{Omar Shah}
\affiliation{Department of Physics and Astronomy, Vrije Universiteit Amsterdam, 1081HV Amsterdam, Netherlands}%
\author{Joachim O. Rädler}
\affiliation{Faculty of Physics and Center for NanoScience, Ludwig-Maximilian-University, Geschwister-Scholl-Platz 1, D-80539 Munich, Germany}
\author{Chase P. Broedersz}%
 \email{c.p.broedersz@vu.nl}
\affiliation{Department of Physics and Astronomy, Vrije Universiteit Amsterdam, 1081HV Amsterdam, Netherlands}%
\affiliation{Arnold Sommerfeld Center for Theoretical Physics and Center for NanoScience, Department of Physics, Ludwig-Maximilian-University Munich, Theresienstraße 37, D-80333 Munich, Germany}

\date{\today}% It is always \today, today,

\begin{abstract}
The migratory dynamics of cells can be influenced by the complex micro-environment through which they move. It remains unclear how the motility machinery of confined cells responds and adapts to their micro-environment. Here, we propose a biophysical mechanism for a geometry-dependent coupling between cellular protrusions and the nucleus that leads to directed migration. We apply our model to geometry-guided cell migration to obtain insights into the origin of directed migration on asymmetric adhesive micro-patterns and the polarization enhancement of cells observed under strong confinement. Remarkably, for cells that can choose between channels of different size, our model predicts an intricate dependence for cellular decision making as a function of the two channel widths, which we confirm experimentally.
\end{abstract}

\maketitle
Cell migration underlies several physiological processes, including tissue development~\cite{franz_cell_2002, weijer_collective_2009}, metastasis~\cite{yamaguchi_cell_2005}, and wound healing~\cite{krawczyk_pattern_1971}. During mesenchymal migration, cells rely on protrusion expansion and contraction to explore their micro-environment~\cite{friedl_proteolytic_2009, ridley_life_2011}. Expansion is driven by actin polymerization, while contractions are generated by the motor protein myosin II interacting with actin filaments. This protrusion-driven process is used, for example, by invading cancer cells to squeeze through pores in the extracellular matrix~\cite{friedl_proteolytic_2009} or to guide neuron growth~\cite{gupton_filopodia_2007}. However, biophysically it is unclear how protrusion-based exploration is influenced by physical cues, such as the geometry or adhesiveness of the micro-environment, posing a challenge to understand confined cell migration.

Several experimental assays focus on specific aspects of physically guided cell migration. These include micro-patterning techniques~\cite{singhvi_engineering_1994, chen_geometric_1997} such as homogeneously coated adhesive lanes~\cite{maiuri_actin_2015, amiri_multistability_2023}, as well as unisotropic substrates where migration is guided by confinement~\cite{bruckner_stochastic_2019, pathak_independent_2012, bruckner_geometry_2022} or directed surface patterning~\cite{comelles_cells_2014, caballero_protrusion_2014, caballero_ratchetaxis_2015, lo_vecchio_collective_2020}. While migration behavior on isotropic surfaces can be understood with microscopic models~\cite{hawkins_pushing_2009, recho_contraction-driven_2013, lavi_deterministic_2016, ron_one-dimensional_2020, sens_stickslip_2020}, migration on unisotropic and confining environments has mostly been studied using phenomenological and data-driven models~\cite{bruckner_geometry_2022, comelles_cells_2014, caballero_protrusion_2014}. To understand confined cell migration from basic biophysical principles, a mechanistic model is needed that describes how physical cues such as geometry affect protrusion growth and cellular decision making.

Here, we construct a mechanistic model for protrusion growth and cell migration on simple confining substrates. The key aspect of our model is how protrusion growth, which directs cell movement, is sensitive to the physical properties of the confining substrate, such as adhesiveness or geometry. We demonstrate and test our model by investigating how cells are guided by various physical cues in frequently employed experimental assays. To show that our model is applicable to unisotropic environments, we consider ``ratchet-like" patterns. The migration bias predicted by our model describes previous experiments~\cite{caballero_protrusion_2014, lo_vecchio_collective_2020}. Subsequently, we consider another central aspect of physiological cell migration: lateral confinement, where our model provides insight into protrusion growth stimulation through confinement~\cite{bruckner_geometry_2022, jain_role_2020}. Finally, we illustrate the generalizability of our model by showing that it describes the impact of protrusion confinement on cellular decision making, which we explore experimentally using cells on micro-patterns. 
 
\begin{figure}[tb]
\includegraphics{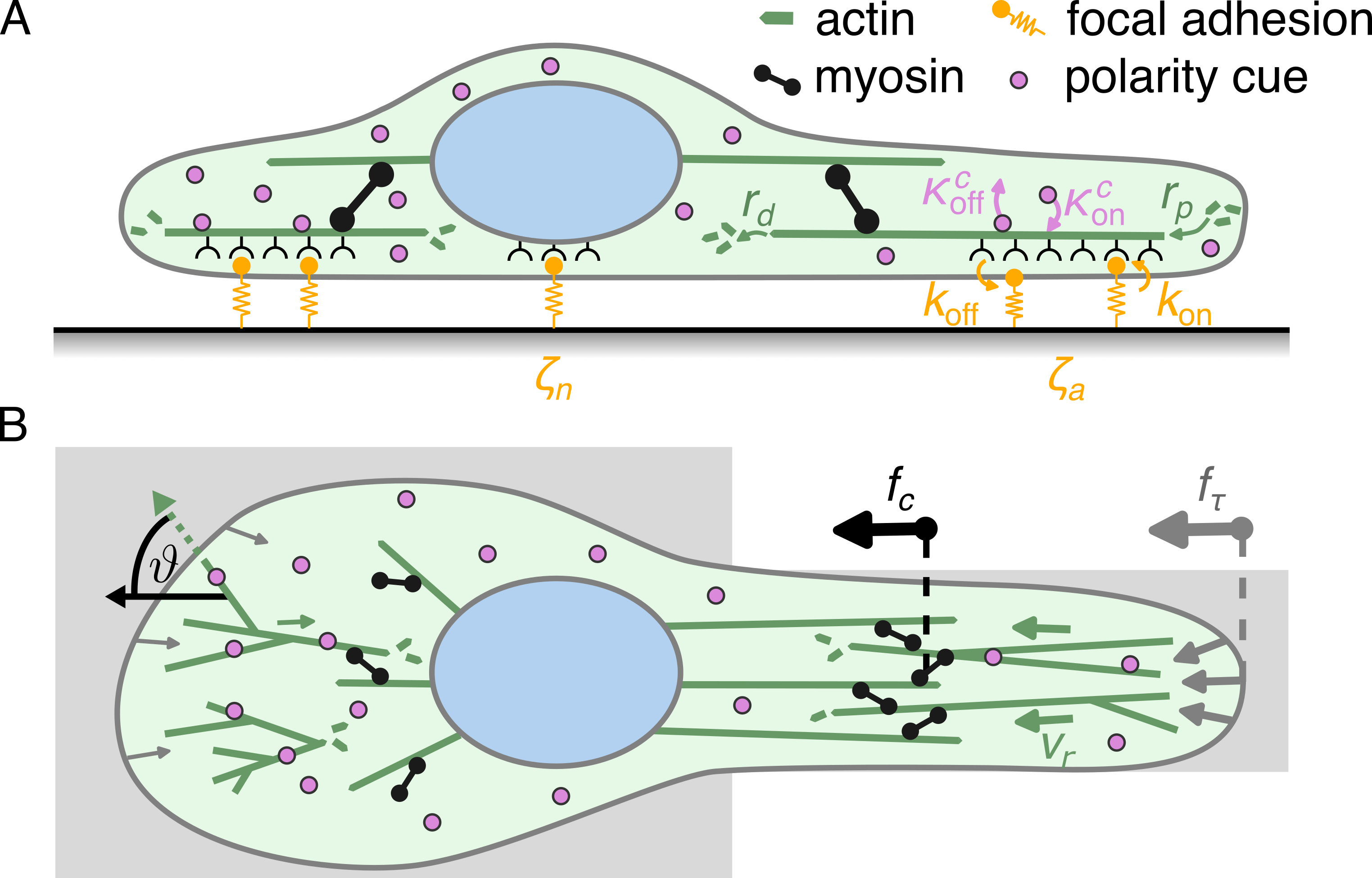}% Here is how to import EPS art
\caption{\label{fig:fig1} \textbf{Components of the migration model.} \textbf{A.} Side view with key molecular components. The stochastic (un)binding of adhesion molecules with rates $k_\mathrm{on}/\mathrm{off}$ gives rise to effective friction coefficients of protrusion ($\zeta_a$) and nucleus ($\zeta_n$). Actin polymerizes at the edge of the protrusions (rate $r_p$) and depolymerizes near the nucleus (rate $r_d$). Polarity cues transiently bind to actin with rates $\kappa_\mathrm{on}^c$ and $\kappa_\mathrm{off}^c$. \textbf{B.} Top view on unisotropic substrate. The retrograde flow (velocity $v_r$) is driven by myosin contractility ($f_c$) and membrane forces ($f_\tau$). Confinement-induced actin alignment (angle $\theta$) stimulates protrusion growth, resulting in increased membrane tension and retrograde flow.}
\end{figure}

We build on models for one-dimensional mesenchymal migration with protrusions on both sides of the nucleus~\cite{lavi_deterministic_2016, ron_one-dimensional_2020, sens_stickslip_2020}. These models account for actin polymerization (rate $r_p$) against a membrane (surface tension $\tau$) (Fig.~\ref{fig:fig1}A). This induces a retrograde flow of actin (velocity $v_\mathrm{r}$) towards the nucleus, which is opposed by focal adhesions connecting actin filaments across the membrane to the substrate. These adhesions can be modeled as elastic bonds that transiently bind to actin filaments, resulting in a friction force~\cite{sens_stickslip_2020}. Based on the observed coupling between retrograde flow velocity and cell polarity~\cite{maiuri_actin_2015}, the retrograde flow is assumed to advect polarity cues, which control actin polymerization rates~\cite{lavi_deterministic_2016, ron_one-dimensional_2020}. Protrusion growth is opposed by myosin contractility and membrane tension. Previously, the resulting restoring force was modeled through the effective material properties of cells, which were assumed to be independent of the environment~\cite{sens_stickslip_2020, ron_one-dimensional_2020}. Hence, the migration dynamics predicted by these models do not couple to the physical micro-environment, restricting their applicability to cell migration on homogeneous substrates.

To broaden the scope of such models, we note that geometry and adhesiveness of the environment determine protrusion shape. To understand cell migration on structured substrates, we derive the protrusion force in terms of its dimensions. We first consider the force balance at the protrusion front, where retrograde flow (velocity $v_r=\zeta_a^{-1}(f_c+f_{\tau})$) is assumed to be driven by the sum of the contractile force $f_c$ and membrane force $f_{\tau}$ (Fig.~\ref{fig:fig1}B), and $\zeta_a$ is the effective friction coefficient due to the binding dynamics of focal adhesions~\footnote{See Supplemental Material at [URL will be inserted by publisher] for details on the model derivation, implementation and parameter values, experimental methods as well as the derivation of an approximate model with a single protrusion coordinate. The Supplemental Material contains Refs.~\cite{evans_forces_2007, bell_models_1978, damiano-guercio_loss_2020, lenz_geometrical_2014, ronceray_fiber_2016, verkhovsky_polarity_1997, iden_crosstalk_2008, erdmann_stochastic_2012, bruckner_disentangling_2020}}. Myosin generates a contractile force by traversing counter-oriented actin filaments inside a network in the crossover region between protrusion and cell body, where actin filaments associated with the nucleus and with adhesions in the protrusion (surface density $\rho_b$) overlap~\cite{svitkina_analysis_1997}. The number of nucleus-associated filaments that contribute to contraction is $N_F^n$. The contractile force is governed by the myosin force-velocity relation~\cite{howard_motilitymodels_2005}, which we approximate here by the stall force $f_s$, since typical nuclear velocities in mesenchymal cell migration ($<30\mathrm{nm/s}$~\cite{bruckner_stochastic_2019, amiri_multistability_2023, gupton_spatiotemporal_2006}) are at least an order of magnitude below the unloaded myosin velocity ($200-800\mathrm{nm/s}$~\cite{howard_speeds_2005}). Considering the influx of actin due to retrograde flow and the loss due to depolymerization (rate $r_d$), the number of actin filaments of diameter $\ell_a$ associated with a protrusion of width $w_p$ is $\rho_bw_p\ell_rv_r/(r_d \ell_a\langle N_b\rangle)$, where $\ell_r$ denotes the width over which focal adhesions are localized and $\langle N_b\rangle$ is the average number of adhesion bonds associated with an actin filament~\footnotemark[\value{footnote}]. A fraction $n_{||}$ of the protrusion associated filaments is aligned with the nucleus-protrusion axis and is thus involved in the protrusion retraction. The myosin number contributing to contractility is then $n_{||}N_F^n\rho_bw_p\ell_rv_r/(r_d \ell_a\langle N_b\rangle) r_m\rho_m$, where $r_m$ is an interaction radius, $\rho_m$ is the myosin line density. Together with the force balance driving retrograde flow, we obtain
\begin{equation}\label{eq:fc_ftau}
    f_c = \tilde{N}_F\rho_bw_p\rho_m f_sv_r = \frac{\tilde{N}_F\rho_b\rho_m w_pf_s}{\zeta_a - \tilde{N}_F\rho_b\rho_mw_p f_s}f_\tau,
\end{equation}
where we introduced the geometry-independent parameter $\tilde{N}_F=n_{||}N_F^nr_m\ell_r/(\ell_ar_d\langle N_b\rangle)$~\footnotemark[\value{footnote}]. Physically, larger membrane forces increase retrograde flow, elevating the actin density in the crossover region and thus increasing contractility (Fig.~\ref{fig:fig1}B). This leads to a coupling between myosin contractility and the membrane force.

To determine the membrane force $f_\tau$, we consider the formation of a protrusion of length $L_p$, height $h_p$, and width $w_p\gg h_p$~\cite{laurent_gradient_2005, shahapure_force_2010}. The resulting increase in surface area is opposed by the membrane force $f_\tau = 2(h_p+w_p)\tau\approx 2w_p\tau$. Note, for constant surface tension $\tau$ the contractile force is independent of protrusion length (Eq.~\eqref{eq:fc_ftau}), which is incompatible with protrusion-guided migration.  However, the surface tension of cells can vary with surface area~\cite{gauthier_temporary_2011, houk_membrane_2012, roffay_passive_2021}. We account for this up to linear order, by $\tau = \tau_0 + 2\tau_1L_p / h_p$, with parameters $\tau_0$ and $\tau_1$, such that
\begin{equation}\label{eq:membrane_force}
f_\tau = 2w_p\tau_0 + \frac{4\tau_1}{h_p}w_pL_p.
\end{equation}
The first term is due to the baseline membrane tension, which should be balanced by the internal cytosolic pressure, such that only the second term contributes to the mechanical coupling between nucleus and protrusion in terms of the contractile force (Eq.~\eqref{eq:fc_ftau})
\begin{equation}\label{eq:contractile_force}
f_c = \frac{4\tau_1\tilde{N}_F\rho_b\rho_m w_pf_s}{\left(\zeta_a - \tilde{N}_F\rho_b\rho_mw_p f_s\right)h_p} w_p L_p = k_c L_p.
\end{equation}
The parameter $k_c$ varies with protrusion position through the local pattern width $w_p(x_{\ell/r})$, where $x_{\ell/r}$ is the position of the left/right protrusion. Note, we defined a linear elastic coupling (spring constant $k_c$) between nucleus and protrusion, which has been used in several migration models~\cite{bruckner_geometry_2022, ron_one-dimensional_2020, sens_stickslip_2020}. In our model, this elastic coupling emerges from the interplay between increased membrane tension, retrograde flow, and myosin contractility. For large forces, load-dependent unbinding kinetics of myosin and adhesions can lead to nonlinearity~\footnotemark[\value{footnote}]. For the experiments considered here, introducing a non-linear coupling between nucleus and protrusion was not needed. In rare cases however, we argue that load-dependent unbinding of myosin is detectable through morphological changes: Unbinding cascades result in failure of rear contractility, which can be described by extending our model (Fig. S2~\footnotemark[\value{footnote}]). 

We combine the result for the contractile force with the polymerization dynamics to determine the protrusion velocity, which is given by the difference between the projected actin polymerization velocity $\ell_a S_{\ell/r} r_p(x_{\ell/r})$ and the retrograde flow velocity, and thus
\begin{equation}\label{eq:xp}
\dot{x}_{\ell/r} = - \frac{k_{\ell/r}(x_{\ell/r})}{\zeta_a(x_{\ell/r})}(x_{\ell/r} - x_n) \mp \ell_aS_{\ell/r} r_p(x_{\ell/r}),
\end{equation}
where we used Eqs.~\eqref{eq:membrane_force}~and~\eqref{eq:contractile_force} and substituted $L_p=x_{\ell/r} - x_n$. The order parameter $S_{\ell/r} = \langle|\cos(\vartheta_{\ell/r})|\rangle$ measures the average orientation of actin filaments (Fig.~\ref{fig:fig1}B) and $k_{\ell/r}(x_{\ell/r}) = k_c(x_{\ell/r}) + 4\tau_1w_p(x_{\ell/r})/h_p$. Since cells can form more adhesions on wider or more densely coated adhesive patterns, $\zeta_a$ can depend on protrusion position~\footnotemark[\value{footnote}]. Eq.~\eqref{eq:xp} illustrates how the confining geometry influences protrusion dynamics: while wider protrusions can produce larger contractile forces (Eq.~\eqref{eq:contractile_force}), they are bound to the substrate through more adhesive bonds. Consequently, on substrates with homogeneous adhesiveness, $k_{\ell/r}/\zeta_a$ is independent of protrusion width and thus identical on both sides of the cell~\footnotemark[\value{footnote}]. This symmetry can be broken due to load-dependent unbinding of adhesions. While this load-sensitivity of adhesion was shown to be key to explain morphologies of moving cells~\cite{sens_stickslip_2020, amiri_multistability_2023}, it was not necessary to include these effects to explain the migration dynamics in the experiments considered here (Fig. S1~\footnotemark[\value{footnote}]).

The nuclear dynamics are given by the balance of the two protrusion forces (Eq.~\eqref{eq:contractile_force}) 
\begin{equation}\label{eq:xn}
\dot{x}_n = \frac{k_c(x_\ell)}{\zeta_n(x_n)}(x_\ell - x_n) + \frac{k_c(x_r)}{\zeta_n(x_n)}(x_r - x_n).
\end{equation}
Since the nucleus is indirectly connected to the substrate through adhesions, the nuclear friction coefficient $\zeta_n$ accounts for internal viscous drag and adhesion-induced stochastic friction~\footnotemark[\value{footnote}].

To complete the model, we require an expression for the polymerization rate $r_p$ (Eq.~\eqref{eq:xp}), which depends on the local polarity cue concentration~\cite{rappel_mechanisms_2017, iden_crosstalk_2008}. As cells polarize, they generate polarity cue gradients. Here, we account for this by considering the concentration of a generic back-polarity cue that accumulates in the cell's rear, inhibiting actin polymerization (Fig.~\ref{fig:fig1}). The difference in average concentration in the two halves of the cell is $\Delta c = c_r - c_\ell$, with $c_{\ell/r}$ representing the concentration to the left/right of the nucleus. The polarity cue binds to and unbinds from actin filaments with rates $\kappa^c_\mathrm{on}$ and $\kappa^c_\mathrm{off}$, respectively. Following~\cite{maiuri_actin_2015}, we assume the polarity cue to be advected with retrograde flow. The advective flux between the two cell parts is then $\Delta v_rn_c c_0$, where $\Delta v_r = v_r(x_r) - v_r(x_\ell)$, $n_c = \kappa^c_\mathrm{on} / (\kappa^c_\mathrm{on} + \kappa^c_\mathrm{off})$ is the bound fraction of polarity cue, and $c_0$ denotes the average cue concentration. The polarity cue flux between the two cell parts is
\begin{equation}\label{eq:flux}
J(x, t) = -(1-n_c)\tilde{D}\partial_x c(x,t) - \Delta v_rn_c c_0 + \tilde{\sigma}\xi(t),
\end{equation}
with diffusion constant $\tilde{D}$. We assume the polarity to be the dominating source of noise in the migratory dynamics of cells, and account for this in Eq.~\eqref{eq:flux} by adding Gaussian white noise $\xi(t)$ of strength $\tilde{\sigma}$.

Treating the average polarity cue concentration $c_0$ as conserved, we write the concentration in the two cell halves as $c_{\ell/r} = c_0  \mp \Delta c/2$ and approximate  $\partial_x c(x,t) \approx \Delta c/ L_c$,  where $L_c$ is the cell length. The dynamics of $\Delta c$ can then be approximated by~\footnotemark[\value{footnote}]
\begin{equation}\label{eq:Deltac}
\partial_t \Delta c(t) \approx -D\Delta c(t) - \frac{4n_c c_0}{L_c} \Delta v_r(\Delta c) +  \frac{4\tilde{\sigma}}{L_c} \xi(t),
\end{equation}
with $D = 4(1-n_c)\tilde{D}L_c^{-2}$. Importantly, for protrusions at steady state, the retrograde flow equals the projected polymerization velocity (see Eq.~\eqref{eq:xp}). Expanding $\Delta v_r$ in terms of $\Delta c$ on both cell sides~\footnotemark[\value{footnote}] yields an equation for the polarity dynamics ($P(t) \equiv -P_0 \Delta c(t)$)
\begin{equation}\label{eq:Polarity}
\dot{P} = - \alpha P - \beta P^3 + \delta \left(S_r - S_\ell\right) + \sigma \xi(t),
\end{equation}
with $\sigma=-4P_0\tilde{\sigma}L_c^{-1}$, $\delta=4P_0n_cc_0L_c^{-1}\ell_a r_p(c_0)$, $\alpha=D-4n_cc_0L_c^{-1}\ell_ar_1\left(S_r+S_\ell\right)$, and $\beta=4P_0^{-2}n_cc_0L_c^{-1}\ell_ar_3(S_r+S_\ell)$.  The coefficients $r_\mathrm{1/3} > 0$ stem from the expansion of $\Delta v_r$. To couple Eq.~\eqref{eq:Polarity} to Eq.~\eqref{eq:xp}, we note that we can relate polarization rate to polarity to leading order as $r_p(t) = r_p(c_0) + r_1 P_0^{-1}P(t)$. The resulting mechanistic model given by the closed set of equations \eqref{eq:xp}, \eqref{eq:xn}, and \eqref{eq:Polarity} describes protrusion-driven 1D cell migration behavior of cells in structured micro-environments. To test this model and investigate what new insights it can give into confined migration, we consider various experimentally studied migration assays.

\begin{figure}[t!]
\includegraphics{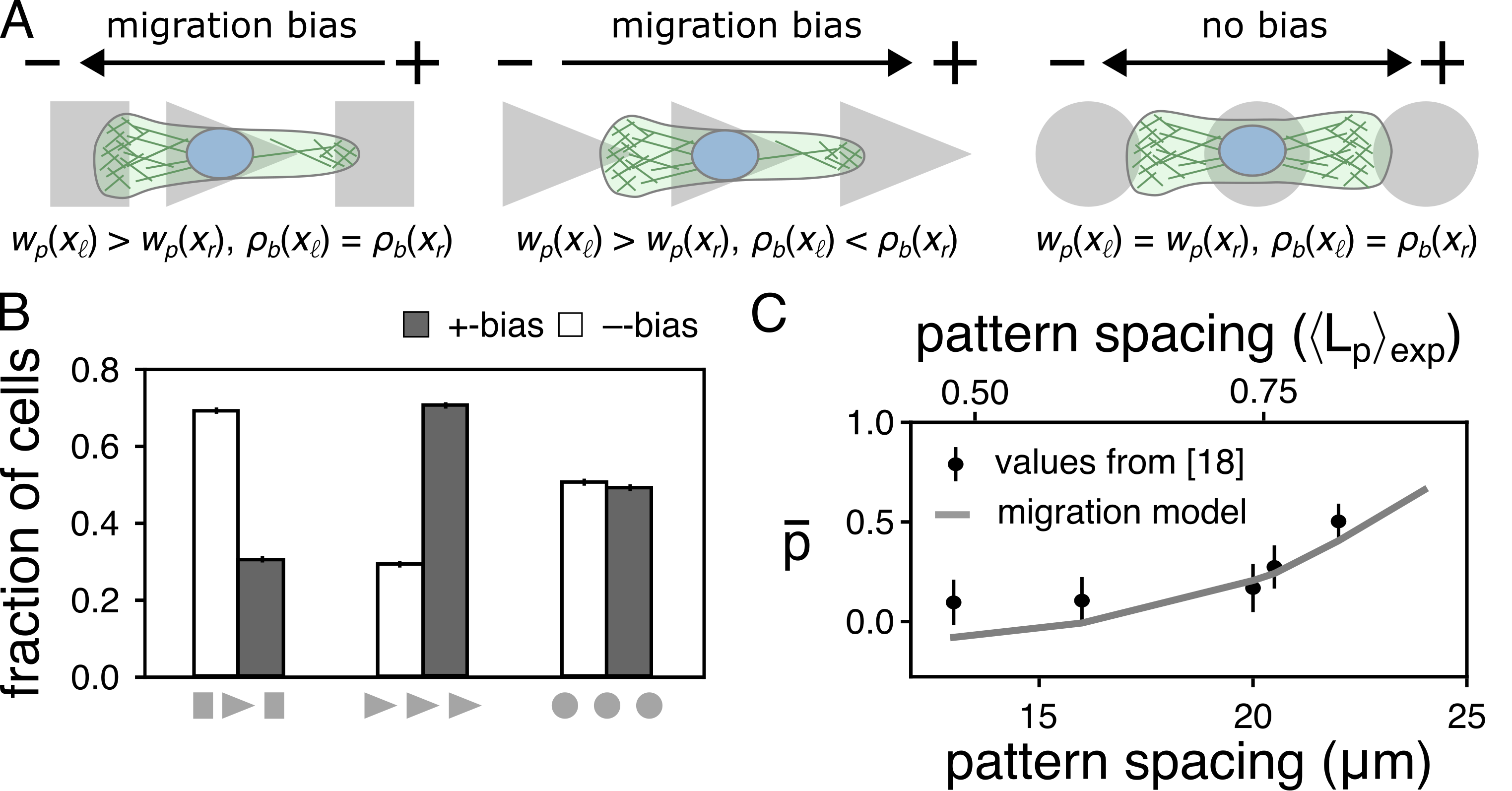}% Here is how to import EPS art
\caption{\label{fig:fig3} \textbf{Cell migration on directed substrates (ratchetaxis).} \textbf{A.} In contrast to symmetric patterns (right), triangular patterns (left, center) lead to protrusions of different widths $w_p$ on both sides of the nucleus. If neighboring patterns are also asymmetric (center) this can result in different densities of adhesive bonds $\rho_b$ at the front of the protrusions. $x_{\ell/r}$ denote the position of the left/right protrusion. These asymmetries in protrusion width and adhesiveness lead to biased migration. \textbf{B.} The model reproduces the experimentally observed first-step migration biases on micro-patterns shown in A. \textbf{C.} Effect of pattern spacing on the average long-term migration bias $\bar{p} = \langle(N_+ - N_-) / (N_+ + N_-)\rangle$ on periodic ratchet-like patterns (A., center), with the number of steps in the $+/-$-direction $N_{+/-}$. $\langle L_p\rangle_\mathrm{exp}$ indicates the average experimentally determined protrusion length (27\textmu m). As observed experimentally, the bias increases with increasing pattern spacing (experimental data form~\cite{lo_vecchio_collective_2020}).}
\end{figure}

\emph{Migration on Directed Patterns.--} Cell migration can be biased due to unisotropic adhesiveness of the micro-environment~\cite{caballero_protrusion_2014, caballero_ratchetaxis_2015, lo_vecchio_collective_2020}. Experimentally, this is realized through ``ratchet-like" adhesive patterns (Fig.~\ref{fig:fig3}A), which induce biased migration of NIH3T3 cells known as ``ratchetaxis". The bias direction depends on micro-pattern geometry. To capture ratchetaxis in our model, we first consider cells on a triangular pattern with two symmetric, rectangular neighboring patterns that the protrusions can engage with (Fig.~\ref{fig:fig3}A, left). The triangular shape allows cells to form wider protrusions on the triangle's blunt end. The geometry sensing mechanism in our model follows from Eq.~\eqref{eq:contractile_force}: Wider protrusions lead to increased spring constants ($w_p$-dependence of $k_c$) and thus $k_c(x_\ell) > k_c(x_r)$~\footnotemark[\value{footnote}]. Since pattern boundaries are not parallel to the migration direction, we assume $S_\ell = S_r$. Hence, the $k_c$ asymmetry leads to higher pulling forces on the nucleus towards the pattern's wider side ($-$-direction), resulting in a migration bias in the $-$-direction consistent with experiments~\cite{caballero_protrusion_2014} (Fig.~\ref{fig:fig3}B). 

On periodic triangular patterns (Fig.~\ref{fig:fig3}A, center) a second mechanism becomes relevant. Due to the tapering of the pattern, protrusions in the $-$-direction overlap with non-adhesive regions. This reduces the adhesion bond density $\rho_b$ and thus $k_\ell > k_r$ ~\footnotemark[\value{footnote}], resulting in a $+$-bias in our model (Fig.~\ref{fig:fig3}B, Fig. S5~\footnotemark[\value{footnote}]). Bigger gaps between patterns lead to stronger asymmetry in the adhesion density below the two protrusions. This reinforces the imbalance between $k_\ell$ and $k_r$ (Fig. S6A~\footnotemark[\value{footnote}]). Together with an observed saturation of adhesive bonds on large adhesion areas, this leads to a non-linear increase of the migration bias that agrees well with experimental data~\cite{lo_vecchio_collective_2020} (Fig.~\ref{fig:fig3}C).

\begin{figure}[t!]
\includegraphics{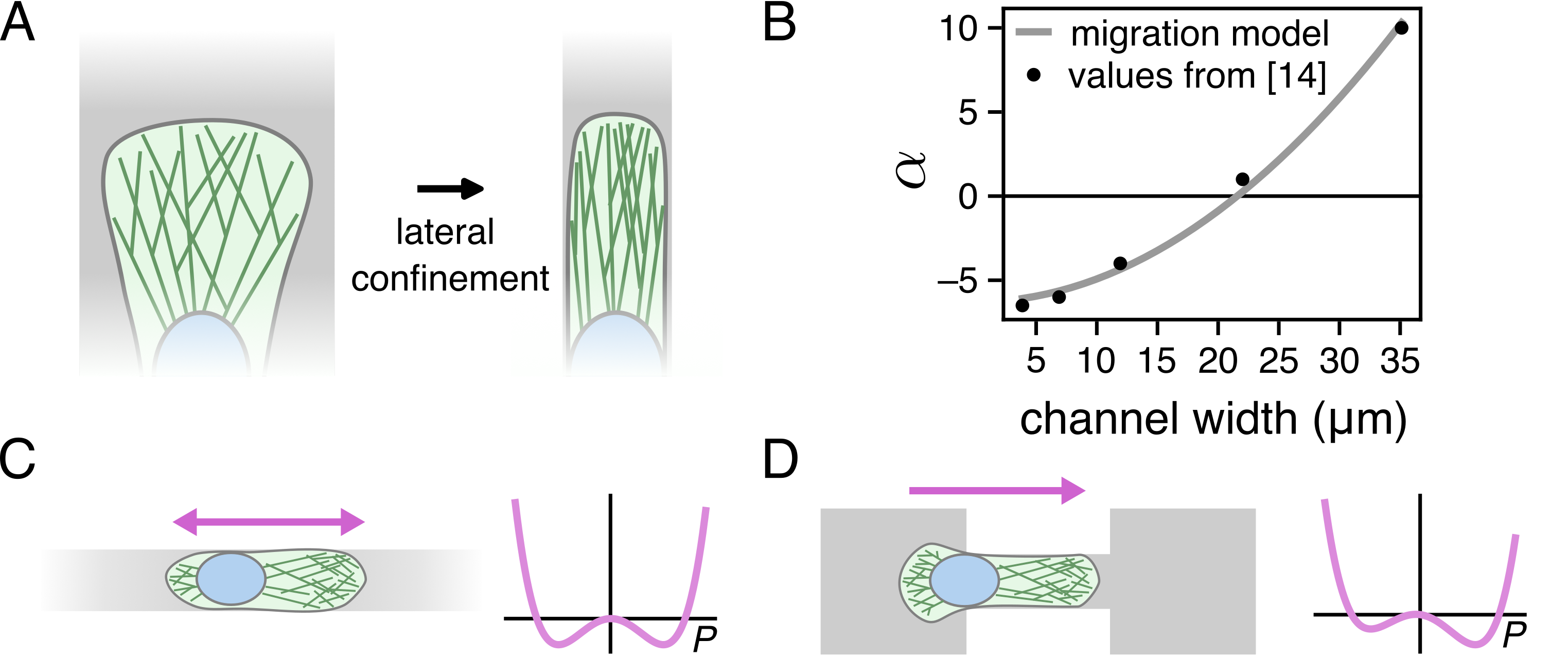}% Here is how to import EPS art
\caption{\label{fig:fig2} \textbf{Cell migration in lateral confinement.} \textbf{A.} Lateral confinement of the protrusion leads to actin filament alignment. \textbf{B.} Fit of expression for $\alpha$ on dumbbell-shaped patterns to the values reported in~\cite{bruckner_geometry_2022}. \textbf{C.} and \textbf{D.} Lateral confinement induces spontaneous polarization of the cell. For homogeneous confinement (C.), both polarization directions are equally likely. For asymmetric confinement (D.), polarization is biased in the direction of stronger confinement.}
\end{figure}

\emph{Lateral Confinement.--} \emph{In vivo}, cells frequently migrate under lateral confinement. Experimentally, the effect of confinement is often studied using micro-patterns~\cite{singhvi_engineering_1994, chen_geometric_1997, sharma_reconstitution_2012, maiuri_actin_2015, schreiber_adhesionvelocity_2021}, and this lateral confinement induces intricate nonlinear migration dynamics~\cite{bruckner_stochastic_2019, bruckner_geometry_2022}. To provide mechanistic insight into these findings, we apply our model to cells migrating through confining adhesive channels. Actin branching~\cite{mullins_interaction_1998} and random fluctuations lead to a range of filament orientations in unconfined cells~\cite{small_actin_1995, svitkina_analysis_1997}.  We expect the distribution of orientations to be reduced by confinement (Fig.~\ref{fig:fig2}A) either due to direct constraints for filaments longer than the pattern width or propagation of a preferred orientation over a correlation length scale into the bulk through alignment interactions, as in liquid crystals~\cite{lee_boundary_1971, saintillan_active_2013, furthauer_self-straining_2019}. 

We account for such boundary-induced alignment by assuming that the order parameter $S$ increases with confinement as $S = 1 - sw_p^2$, where $s>0$ is a parameter that accounts for the strength of actin alignment interactions. This gives us an expression for $\alpha$ (Eq.~\eqref{eq:Polarity}) as a function of protrusion width:
\begin{equation}
\alpha(w_{p, \ell}, w_{p, r}) = D - 4n_c c_0L_c^{-1} l_a r_1(2 - sw_{p, r}^2 - sw_{p, \ell}^2).
\end{equation}
When protrusions are sufficiently confined, $\alpha<0$ inducing a self-reinforcement of the polarity and consequently protrusion growth. For $\delta > 0$, increased actin alignment on one side of the cell results in a finite polarization, even if $\alpha > 0$; the larger projected polymerization velocity driving protrusion growth increases retrograde flow on the confined side and thus accumulates a concentration gradient. This effect also biases polarization for $\alpha < 0$ towards the cell's confined side. 

The polarity feedback parameter $\alpha(w_{\rm cell}, w_c)$ for unconfined cells (width $w_{\rm cell}$) entering a confining channel (width $w_c$) with their leading protrusion (Fig.~\ref{fig:fig1}B) is shown in Fig.~\ref{fig:fig2}B, and agrees well with a similar, inferred model constrained by experimental data~\cite{bruckner_geometry_2022}. In uniform confinement, both directions of polarization are equally likely (Fig.~\ref{fig:fig2}C). However, when only one side of the cell is confined (Fig.~\ref{fig:fig2}D), our model predicts a bias for cells to polarize towards confinement. For migration on dumbbell-shaped micro-patterns we find the same qualitative behavior reported in~\cite{bruckner_geometry_2022}~\footnotemark[\value{footnote}]. The model developed here offers a biophysical mechanism underlying the geometry adaptation of protrusion and polarity dynamics reported previously: the sign of $\alpha(w_{p, \ell}, w_{p, r})$ is set by the competition between advection and diffusion of polarity cues. Confinement-induced protrusion growth increases the advection of polarity cue towards the cell rear. As a consequence, differences in the polarity concentration get reinforced under confinement.

\begin{figure}[t!]
\includegraphics{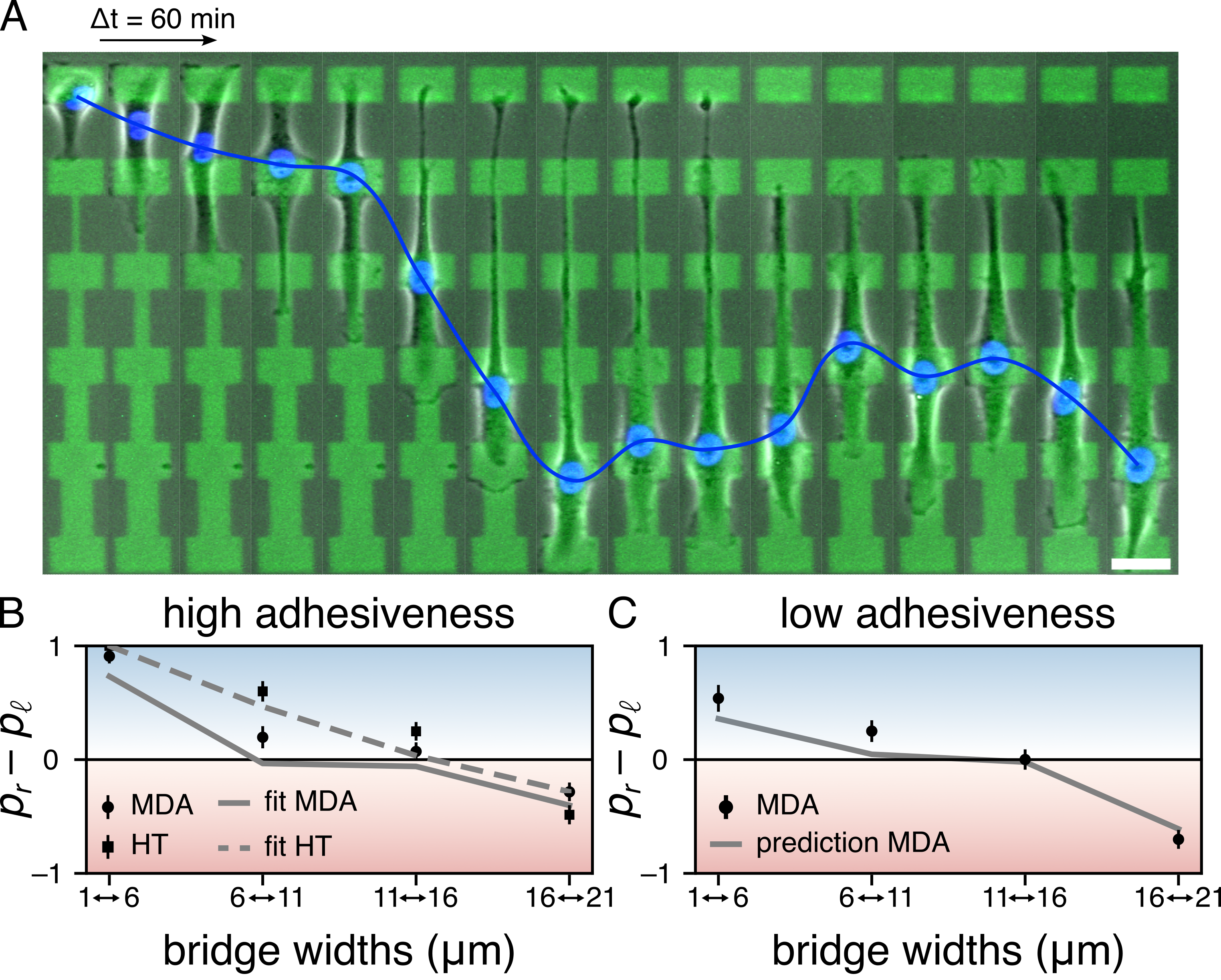}% Here is how to import EPS art
\caption{\label{fig:fig4} \textbf{Cellular decision making in lateral confinement.} \textbf{A.} Time series of a cell migrating on a chain of adhesive islands that are connected by adhesive bridges of increasing width (scale bar: 35\textmu m). \textbf{B.} Experimentally observed migration biases together with model fits for MDA-MB-231 and HT-1080 cells, where $p_r/\ell$ denotes the probability for a cell to choose the right/left channel. With increasing bridge widths, cells transition from a bias towards wider bridges to a bias towards narrower bridges. \textbf{C.} Predicted migration biases of MDA-MB-231 cells on patterns of reduced adhesiveness together with experimental data.}
\end{figure}

Our model can be generalized to other confining geometries, as opposed to prior approaches~\cite{caballero_protrusion_2014, sens_stickslip_2020, bruckner_geometry_2022}. In particular, based on our model, we expect that confinement can have opposing effects on the preferred direction of migration. On the one hand, confinement-induced actin alignment stimulates polarization towards reduced channel width. On the other hand, contractile forces decrease with confinement (Eq.~\eqref{eq:contractile_force}), leading to a net force on the nucleus away from confinement. For concreteness, consider cells that can choose between two channels of different width. Our model predicts that the preferred migration direction depends on the two channel widths and the overall cellular contractility and polarization~\footnotemark[\value{footnote}]. In particular, we expect a transition from favoring wider to narrower channels with increasing width (Fig. S3A~\footnotemark[\value{footnote}]) and that with increasing contractility and polarization narrower channels are favored. To test this, we perform micro-pattern experiments with two different mesenchymal cell lines (MDA-MB-231 and HT-1080) migrating on a series of square-shaped adhesive islands connected by a range of bridges with increasing width (Fig.~\ref{fig:fig4}A)~\footnotemark[\value{footnote}]. Our experiments confirm the expected transition of migration biases with increasing channel width and this trend is quantitatively captured by our model (Fig.~\ref{fig:fig4}B). The parameters underlying the model fits suggest that HT-1080 cells are less contractile and less polarized than MDA-MB-231 cells (Fig.~S6~\footnotemark[\value{footnote}]). To explore how cellular behavior depends on experimental conditions and to further test our model, we repeat the experiment with MDA-MB-231 cells on patterns with 50\% reduced adhesiveness. Notably, after all model parameters are fixed by fitting experiments on high-adhesion patterns, we can quantitatively predict the observed changes in migration biases (Fig.~\ref{fig:fig4}C). The observed transition in migration biases yields new insights into cellular decision making in confining environments with heterogeneous pore sizes. 

To summarize, we developed a generalizable model for directed mesenchymal cell migration in structured micro-environments from basic biophysical principles. At the core of our model is the coupling between substrate-controlled protrusion formation and cellular migration behavior. Here, it would be insightful to further investigate how the micro-environment affects membrane tension locally and globally. We demonstrated that our model explains the emergence of directed migration of a number of different cell lines in response to different external cues such as asymmetric adhesion densities and the geometry sensing of polarity dynamics~\cite{bruckner_geometry_2022} in lateral confinement. This model of confined cell migration broadens the scope of cell migration models to more physiological conditions, where cells are simultaneously exposed to different physical migration cues and could form the basis for more complex approaches in the future that also integrate detailed biochemical signaling pathways.

We thank Daniel Riveline, Tom Brandstätter, Janni Harju, David Brückner and Bram Hoogland for helpful discussions. This project was funded by the Deutsche Forschungsgemeinschaft (DFG, German Research Foundation), Project No. 201269156— SFB 1032 (Projects B01 and B12). 

\bibliography{references}% Produces the bibliography via BibTeX.

\end{document}

\renewcommand{\thefigure}{S\arabic{figure}}

\title{Geometry-sensitive protrusion growth directs confined cell migration}% Force line breaks with \\

\author{Johannes Flommersfeld}
\affiliation{Department of Physics and Astronomy, Vrije Universiteit Amsterdam, 1081HV Amsterdam, Netherlands}%
\affiliation{Arnold Sommerfeld Center for Theoretical Physics and Center for NanoScience, Department of Physics, Ludwig-Maximilian-University Munich, Theresienstraße 37, D-80333 Munich, Germany}
\author{Stefan Stöberl}
\affiliation{Faculty of Physics and Center for NanoScience, Ludwig-Maximilian-University, Geschwister-Scholl-Platz 1, D-80539 Munich, Germany}
\author{Omar Shah}
\affiliation{Department of Physics and Astronomy, Vrije Universiteit Amsterdam, 1081HV Amsterdam, Netherlands}%
\author{Joachim O. Rädler}
\affiliation{Faculty of Physics and Center for NanoScience, Ludwig-Maximilian-University, Geschwister-Scholl-Platz 1, D-80539 Munich, Germany}
\author{Chase P. Broedersz}%
 \email{c.p.broedersz@vu.nl}
\affiliation{Department of Physics and Astronomy, Vrije Universiteit Amsterdam, 1081HV Amsterdam, Netherlands}%
\affiliation{Arnold Sommerfeld Center for Theoretical Physics and Center for NanoScience, Department of Physics, Ludwig-Maximilian-University Munich, Theresienstraße 37, D-80333 Munich, Germany}

\maketitle

\usetagform{supplementary}
\section{Details of the model derivation}

\subsection{Internal and stochastic friction}
We consider the friction forces that act on the protrusions and the nucleus. There are two contributions to this friction: internal friction due to the interaction of actin filaments/the nucleus with the cytosol and stochastic friction due to the binding and unbinding dynamics of the focal adhesions. We first consider the friction forces acting on the actin filaments at the front of the protrusion $x_f$. The total force exerted by the adhesion bonds is given by $f_\mathrm{ad} = \rho_b\langle n f_b\rangle \ell_r w_p$, where $\rho_b$ is the total surface density of bonds, $n$ is the fraction of bound bonds, $f_b$ is the force per bond, $w_p$ is the width of the protrusion and $\ell_r$ is the region at the front of the protrusion over which the retrograde flow is concentrated. When approximating $\langle n f_b \rangle \approx n \langle f_b\rangle$, one can express the total force exerted by bound adhesion bonds in terms of the retrograde flow as~\cite{sens_stickslip_2020} 
\begin{equation}
f_\mathrm{ad} = \zeta_1^a (f_b) v_r
\end{equation}
 with 
 \begin{equation}\label{eq:friction_coefficitient_adhesions}
 \zeta_1^a(f_b) = \frac{\rho_b \ell_r k_b n(f_b)w_p}{k_\mathrm{off}(f_b)}.
\end{equation} 
Since the binding and unbinding dynamics of the bonds can be mechanosensitive, $\zeta_1^a$ will in general depend on the force per bond $f_b$. Overall, we see that the collective binding and unbinding dynamics of the adhesion bonds lead to a friction force acting on the F-actin in the protrusion.

Additional to this stochastic friction, there is also internal friction that the actin filaments experience as they move through the cytosol. We write this internal friction as $f_i^a = \zeta_0^a v_r$, where $\zeta_0^a$ is the internal friction coefficient of actin filaments. Taken together we get the following expression for the total friction force:
\begin{equation}\label{eq:actin_friction}
f_f^a = \left(\zeta_0^a + \zeta_1^a(f_b)\right)v_r,
\end{equation}

Analogously, a similar friction force acts on the nucleus, which we write as 
\begin{equation}\label{eq:nuclear_friction}
f_f^n = \left(\zeta_0^n + \zeta_1^n(f_b))\right)v_n.
\end{equation}
Here, we replaced the retrograde flow with the nuclear velocity $v_n$ and  $\zeta_1^n(f_b) = \rho_b \ell_n k_b n(f_b) w_n / k_\mathrm{off}(f_b)$, where $\ell_n$ and $w_n$ denote the length and the width over which the nucleus is mechanically interacting with focal adhesions.

 In the case of purely surface-confined cell migration, it was found that it is not necessary to account for the mechanosensitivity of the adhesion bonds to explain the observed migration behavior~\cite{sens_stickslip_2020, bruckner_geometry_2022}. Hence, we assume that the load per adhesion bond $|f_b|$ is negligible compared the characteristic molecular force scale $f_b^*$~\cite{evans_forces_2007, bell_models_1978} and thus $n = n_0$ and $k_\mathrm{off} = k_\mathrm{off}^0$. Furthermore, for the friction acting on the front of the cell, we expect the focal adhesions to be the main contribution to the total friction due to the importance of focal adhesions for the formation of intact lamellipodia~\cite{damiano-guercio_loss_2020}. Consequently, we can write the total friction acting on the front of the cells as
\begin{equation}\label{eq:protrusion_friction}
f_f^a \approx \zeta_1^a(0) v_\mathrm{r} = \zeta_a v_r,
\end{equation}
where we defined $\zeta_a = \zeta_1^a(0) = \rho_b \ell_r k_b n_0 w_p / k_\mathrm{off}^0$. 

In contrast, since the nucleus is much larger than the actin filaments, it is less clear if the internal friction can be neglected. Hence, we keep both sources of friction in that case and get
\begin{equation}\label{eq:nuclear_friction_low_force}
f_f^n = \left(\zeta_0^n + \zeta_1^n(0)\right)v_n = \zeta_n v_n.
\end{equation}

\subsection{Actin density and orientation in the crossover region}
In the crossover region between the protrusion and the nucleus, new filamentous actin is transported in with a rate that is proportional to $v_r / \ell_a$, where $\ell_a$ denotes the size of an actin monomer. To transmit myosin-generated tension between the nucleus and the front of the protrusions, filaments need to be anchored to adhesions at the front of the protrusion. The number of protrusion-associated filaments than can contribute to the contraction is thus proportional to the number of adhesions in the front of the protrusion $\rho_b w_p \ell_r$. The total rate of influx of protrusion-associated actin is thus given by $\rho_b w_p \ell_r \langle N_b\rangle^{-1} v_r/\ell_a$, where $\langle N_b \rangle$ denotes the average number of adhesive bonds per actin filament. Towards the nucleus, we expect the retrograde flow to cease such that the main contribution to the loss of filamentous actin in that region is actin depolymerization with a constant rate $r_d$. The number of protrusion-associated actin filaments $N_F^{p}$ in the crossover region is then given by
\begin{equation}
\dot{N}_F^p(t) = \frac{v_r \rho_b w_p \ell_r}{\ell_a\langle N_b\rangle} - r_dN_F^p(t).
\end{equation} 
At steady state we thus get 
\begin{equation}\label{eq:nF}
N_F^p(v_r) = \frac{v_r  \rho_b w_p \ell_r}{r_d\ell_a\langle N_b\rangle}.
\end{equation}

In the crossover region, myosin links actin filaments coming from the front of the protrusion to filaments associated with the nucleus~\cite{svitkina_analysis_1997}. When a myosin motor engages with two actin filaments from these two respective networks, their relative orientation dictates if the generated local force is contractile~\cite{lenz_geometrical_2014}. To account for this, we distinguish between three qualitatively different classes of filament orientations: filaments oriented with their barbed ends towards the front of the protrusion, filaments oriented with their pointed ends towards the front of the protrusion, and filaments oriented orthogonal to the direction of migration. Near the tip of the protrusion, the vast majority of actin filaments is oriented with their barbed ends towards the front of the protrusion (close to $80\%$)~\cite{small_actin_1995, svitkina_analysis_1997}. A smaller fraction of actin filaments (about $20\%$) is oriented orthogonal to the direction of migration and thus not actively contributing to contraction and only a negligible number of filaments is oriented with their pointed end towards the front of the protrusion. As a consequence, the retrograde flow predominantly transports actin filaments into the crossover region that are oriented with their barbed ends towards the protrusion. In the contractile region between the protrusion and the cell body, the fraction of orthogonally oriented filaments increases, likely due to actin reorientation caused by myosin contractility~\cite{svitkina_analysis_1997}.

In the crossover region, the incoming protrusion-associated actin filaments are connected by myosin motors to an actin network of mixed orientation~\cite{verkhovsky_polarity_1997}. This leads to mix of relative orientations of the connected filaments, resulting in different force dipoles: Approximately anti-parallel filaments will generate a force dipole along the nucleus-protrusion axis. Due to the dominant fraction of filaments oriented with their barbed ends towards the protrusion front compared to filaments with the opposite orientation~\cite{small_actin_1995, svitkina_analysis_1997}, we expect the resulting force dipoles to be almost exclusively contractile~\cite{lenz_geometrical_2014}. The imbalance between contractile and extensile stresses is likely to be further enhanced due to buckling of actin filaments~\cite{ronceray_fiber_2016}. Approximately parallel filaments on the other hand, will either result in no force at all or a contractile force dipole orthogonal to the nucleus-protrusion axis~\cite{lenz_geometrical_2014}. These configurations will thus not contribute to the force transmission between the front of the protrusion and the nucleus. The overall contractile force generated in the crossover region will thus depend on the number of incoming filaments that are oriented along the nucleus-protrusion axis and the number of favorably oriented filaments in the disordered network associated with the nucleus $N_F^n$. Due to the disordered nature of the nucleus associated network, we expect the latter to be approximately constant. 

The number of myosin motors that can interact with an actin filament is given by $r_m \rho_m$, where $r_m$ denotes the myosin interaction radius and $\rho_m$ is the myosin line density. Thus, the number of contractile actomyosin configurations $N_\mathrm{am}$ is then given by
\begin{equation}\label{eq:Na}
N_\mathrm{am} = n_{||}r_m\rho_m N_F^{n}N_F^{p}(v_r) = \tilde{N}_F\rho_b\rho_m w_p v_r,
\end{equation}
where we introduced the geometry-independent parameter $\tilde{N}_F = n_{||}N_F^nr_m\ell_r /(\ell_a r_d\langle N_b \rangle)$, with $n_{||}$ the fraction of incoming actin filaments oriented parallel to the nucleus-protrusion axis. Multiplying Eq.~\eqref{eq:Na} by the stall force of a single myosin filament $f_s$ leads to Eq.~(1) in the main text.

\subsection{Candidates for a back-polarity cue}
Our model does not rely on specific properties of the polarity cue other than that it is advected with the actin retrograde flow and that it is increased in the rear of a polarized cell. As a consequence, there is a number of possible candidates for such a polarity cue. 

A previously considered candidate for an actin advected polarity cue is the small GTPase Cdc42. Experimental inhibition of Cdc42, however, showed that it cannot play the role of an advective polarity cue~\cite{maiuri_actin_2015}. Additionally, Cdc42 would not fit our requirements as it is enriched at the front of a polarized cell. A natural candidate that does satisfy our requirements would by myosin due to its localization in the rear of a polarized cell and its involvement in symmetry breaking during cell polarization~\cite{yam_actinmyosin_2007}. This is further supported by experimental work that showed that inhibition of myosin disrupts the correlation between retrograde flow velocity and cell polarity~\cite{maiuri_actin_2015} and theoretical work that showed that myosin II as a back polarity marker can explain the switching between different migration patterns in dendritic cells~\cite{lavi_deterministic_2016}. This makes myosin II the most natural candidate for a general back polarity cue. Other candidates, including other members of the small GTPases that are associated with the rear of the cell such as RhoA~\cite{iden_crosstalk_2008}, can also not be excluded and further experimental characterization will be needed to connect the general concept of an actin advected back-polarity cue to concrete biological polarity markers.

\subsection{Approximation of the advection-diffusion equation with no-flux boundaries}
Starting from Eq.~(6) in the main text, we approximate $\partial_x c\approx \Delta c / L_c$. From this, we get
\begin{equation}\label{eq:flux}
    J(\Delta c, t) \approx -(1-n_c)\tilde{D}L_c^{-1}\Delta c - n_c c_0 \Delta v_r +\tilde{\sigma} \xi(t).
\end{equation}
The local change in concentration of the polarity cue is given by 
\begin{equation}\label{eq:continuity}
    \frac{\partial c(x, t)}{\partial t} = -\frac{\partial J(x, t)}{\partial x}.
\end{equation}
In principle Eq.~\eqref{eq:continuity} describes the concentration profile along the entire long axis of the cell. We are, however, only interested in the average difference between the side edges of the cell. We treat the membrane as a no-flux boundary, such that the only flux of polarity cue in or out of the two halves of the cell is through the midplane. Discretizing the gradient of the flux for each half of the cell then gives
\begin{equation}
    \frac{\partial c_{r}(t)}{\partial t} \approx -\frac{0 - J}{L_c/2} =  2L_c^{-1}J
\end{equation}
and
\begin{equation}
    \frac{\partial c_{\ell}(t)}{\partial t} \approx -\frac{J-0}{L_c/2} = -2L_c^{-1}J.
\end{equation}
Using $\Delta c = c_r - c_\ell$, we then get an approximate expression for the dynamics of $\Delta c$ in a cell with no-flux boundary conditions 
\begin{equation}\label{eq:DeltaC}
    \frac{\partial \Delta c}{\partial t} \approx \frac{4J(\Delta c, t)}{L_c}.
\end{equation}
Plugging Eq.~\eqref{eq:flux} into Eq.~\eqref{eq:DeltaC} then yields Eq.~(7) from the main text.

\subsection{Expansion of the retrograde flow velocity}
To find a simple expression for the difference in retrograde flow velocities $\Delta v_r(\Delta c) = \ell_a S_r r_\mathrm{p}(x_r) - \ell_aS_\ell r_p(x_\ell)$, we split $r_p$ in an even ($r_p'$) and an odd ($r_p''$) part in $\Delta c$. Together with $c(x_{\ell/r}) = c_0 \mp \Delta c(t)/2$, we can then write 
\begin{equation}\label{eq:expansion1}
\Delta v_r (\Delta c) = \ell_a r_p'(\Delta c)(S_r - S_\ell) + \ell_a r_p''(\Delta c)(S_r + S_\ell),
\end{equation}
The even term in $\Delta c$ will enter in Eq.~(7) from the main text as a bias term, which dilutes the side of the cell in which the actin filaments are more aligned with the direction of migration. Due to its simple effect as a bias term, we do not expect the detailed functional dependence of $r_p'$ to be of great effect and thus only keep the leading order term such that $r_p'\approx r_p(c_0)$. The odd term, however, will either dampen or reinforce the concentration gradient depending on its sign. Since we are considering a back-polarity cue, we expect that the odd term leading order has a reinforcing effect on the concentration gradient. To ensure that $\Delta c$ remains bound, we expand $r_p''$ up to third order, such that $r_p''(\Delta c) \approx - r_1 \Delta c + r_{3} \Delta c^3$, with $r_{1/3}>0$. This gives the following approximate expression for $\Delta v_r$:
\begin{equation}\label{eq:expansion2}
\Delta v_r(\Delta c) = \ell_a r_p(c_0)(S_r - S_\ell) - \ell_a r_1(S_r + S_\ell) \Delta c + \ell_a r_3(S_r + S_\ell) \Delta c^3
\end{equation}
Note that the first term on the right-hand side of Eq.~\eqref{eq:expansion2} leads to a difference in retrograde flows even in the case of a homogeneous concentration profile of the polarity cue. This difference is of purely geometric origin: If the actin filaments on one side of the cell are more aligned, their projected growth velocity $l_ar_pS_{\ell/r}$ is larger at the same polymerization rate $r_p$. This leads to faster protrusion growth on that side and due to the coupling between protrusion length and retrograde flow velocity (see Eqs.~(1)~and~(3), main text), this results in a faster retrograde flow at steady state. Over time, this will build up a concentration gradient in the cell according to Eq.~(7) in the main text, with a lower concentration of the back polarity cue on the side with stronger alignment. The negative effect of the polarity cue on the polymerization rate further reinforces the imbalance in protrusion growth and thus retrograde flow velocities between the two sides. If the alignment is strong enough and in the absence of other external cues, this leads to a positive feedback loop that results in a polarization of the cell towards the side of the cell with higher actin alignment.

\subsection{Effects of force-dependent unbinding kinetics}
So far, we have neglected the load-dependence of the unbinding kinetics of both myosin motors and adhesion bonds. Here, we extend our model beyond this limit to demonstrate the generality of our model and the conclusions in the main text. We first derive the leading-order correction to the model to get an intuition of the effect that load-dependence of the unbinding kinetics would have on the model before qualitatively discussing the effect of possible unbinding cascades.

The force sensitivity of the myosin unbinding kinetics is reflected in its force-velocity relation~\cite{erdmann_stochastic_2012}. In the main text, we approximated the force that a single myosin motor generates by the stall force. To go beyond this approximation, we linearize the myosin force-velocity relation around the stall force, such that the force $f_m$ that a single myosin filament generates is given by
\begin{equation}\label{eq:force_velo1}
f_m \approx f_s - c_m v_r,
\end{equation}
where $c_m$ is a positive constant. Since the total contractile force $f_c$ is to leading order proportional to the retrograde flow velocity (Eq.~(1), main text), we can write $f_m$ in terms of the total contractile force to leading order as
\begin{equation}\label{eq:force_velo2}
f_m(f_c) \approx f_s -  \frac{c_m}{\tilde{N}_F w_p \rho_b \rho_m f_s}f_c = f_s - \tilde{c}_mf_c.
\end{equation} 
Hence with increasing overall contractility, the force that an individual myosin motor can generate decreases.

Next, we consider the unbinding kinetics of the adhesion bonds. These give rise to force-sensitive friction coefficients $\zeta_a(f_b)$ and $\zeta_n(f_b)$ (see Eqs.~\eqref{eq:actin_friction}~and~\eqref{eq:nuclear_friction}). We can then expand the stochastic contribution $\zeta_1^{a/n}$ to the friction coefficient to leading order in the average force per bond $f_b$ such that
\begin{equation}
\zeta_1^{a/n} \approx \zeta_1^{a/n}(0) + c_\zeta^{a/n}f_b,
\end{equation}
where the sign of $c_\zeta^{a/n}$ determines wheter the adhesive bonds act as catch ($c_\zeta^{a/n}>0$) or slip ($c_\zeta^{a/n}<0$) bonds. Using that we can express the average load per adhesion bonds as $f_b=k_bv_{r/n}/k_\mathrm{off}(f_b)$~\cite{sens_stickslip_2020}, we get the following leading order approximation of the stochastic friction coefficients:
\begin{equation}
\zeta_1^{a/n} \approx \zeta_1^{a/n}(0) + c_\zeta^{a/n}\frac{k_b}{k_\mathrm{off}(0)}v_{r/n}.
\end{equation}
Due to the size of the nucleus, we expect its viscous drag to play a significant role in the overall nuclear friction coefficient, limiting the achievable nuclear velocities. As a consequence, in the following analysis, we focus on the force dependence of $\zeta_1^{a}$ and keep $\zeta_1^{n}$ constant. Using Eq.~(1), we can then relate $\zeta^{a}$ to the contractile force to leading order through 
\begin{equation}\label{eq:fric_correction}
\zeta^a(f_c) \approx \zeta_a(0) + \tilde{c}_\zeta^a\frac{k_b}{k_\mathrm{off}(0)}f_c,
\end{equation}
where $\tilde{c}_\zeta^a$ is defined analogously to $\tilde{c}_m$ in Eq.~\eqref{eq:force_velo2}.

Plugging Eqs.~\eqref{eq:force_velo2}~and~\eqref{eq:fric_correction} into Eq.~(3) from the main text, we get the following leading order correction to the linear coupling between the nucleus and the protrusion
\begin{equation}\label{eq:fc_corr1}
f_c \approx \frac{4h_p^{-1}\tau_1 \tilde{N}_Fw_p^2 \rho_b\rho_m (f_s-\tilde{c}_mf_c)}{\zeta_a(0) + \tilde{c}_\zeta^ak_bf_c/k_\mathrm{off}(0) - \tilde{N}_Fw_p\rho_b\rho_m (f_s-\tilde{c}_mf_c)} L_p.
\end{equation}
Expanding Eq.~\eqref{eq:fc_corr1} leads to 
\begin{equation}\label{eq:fc_corr2}
f_c \approx k_c(0) L_p - (k_1\tilde{c}_m+k_2\tilde{c}_\zeta^a)L_p^2,
\end{equation}
with
\begin{equation}
k_1 = \frac{k_c(0)^2}{f_s}\left(1 + \frac{k_c(0)h_p}{4\tau_1w_p}\right)>0
\end{equation}
and
\begin{equation}
k_2 = k_c(0)^2\frac{k_bk_\mathrm{off}(0)^{-1}}{\zeta_a(0)-\tilde{N}_F(0)w_p\rho_b\rho_mf_s}>0.
\end{equation}

Eqs.~\eqref{eq:fric_correction}~and~\eqref{eq:fc_corr2} allow us to develop an intuition for the qualitative effects that force-sensitive bonds have on migration behavior. The load-sensitive unbinding kinetics of myosin II weakens the coupling between nucleus and protrusion. This is particularly pronounced at narrow channel widths, where $k_1/k_c(0)$ increases due to the channel width dependence of $\tilde{c}_\zeta^a$. This comes on top of the weakening of the contractile force due to the width dependence of $k_c(0)$  that is discussed in the main text (see Eq.~(3), main text). The effect that the unbinding kinetics of adhesion bonds have on the model depends on the sign of $\tilde{c}_\zeta^a$. In the case of catch bonds, $k_1 \tilde{c}_m+k_2\tilde{c}_\zeta^a > 0$ and the load-dependence of the adhesion unbinding kinetics further adds to the weakening of the coupling between nucleus and protrusion. In combination with the increased friction of the protrusions (Eq.~\eqref{eq:fric_correction}), this would result in cells displaying an elongated morphology to the narrower side when migrating on asymmetric patterns. In the case of slip bonds, $k_1\tilde{c}_m+k_2\tilde{c}_\zeta^a$ would decrease or even become negative, resulting in a stronger coupling between the protrusion and the nucleus. Together with the decreased friction of the protrusions, this would reduce the length and stability of protrusions that can be formed. 

\begin{figure*}[htb]
\includegraphics{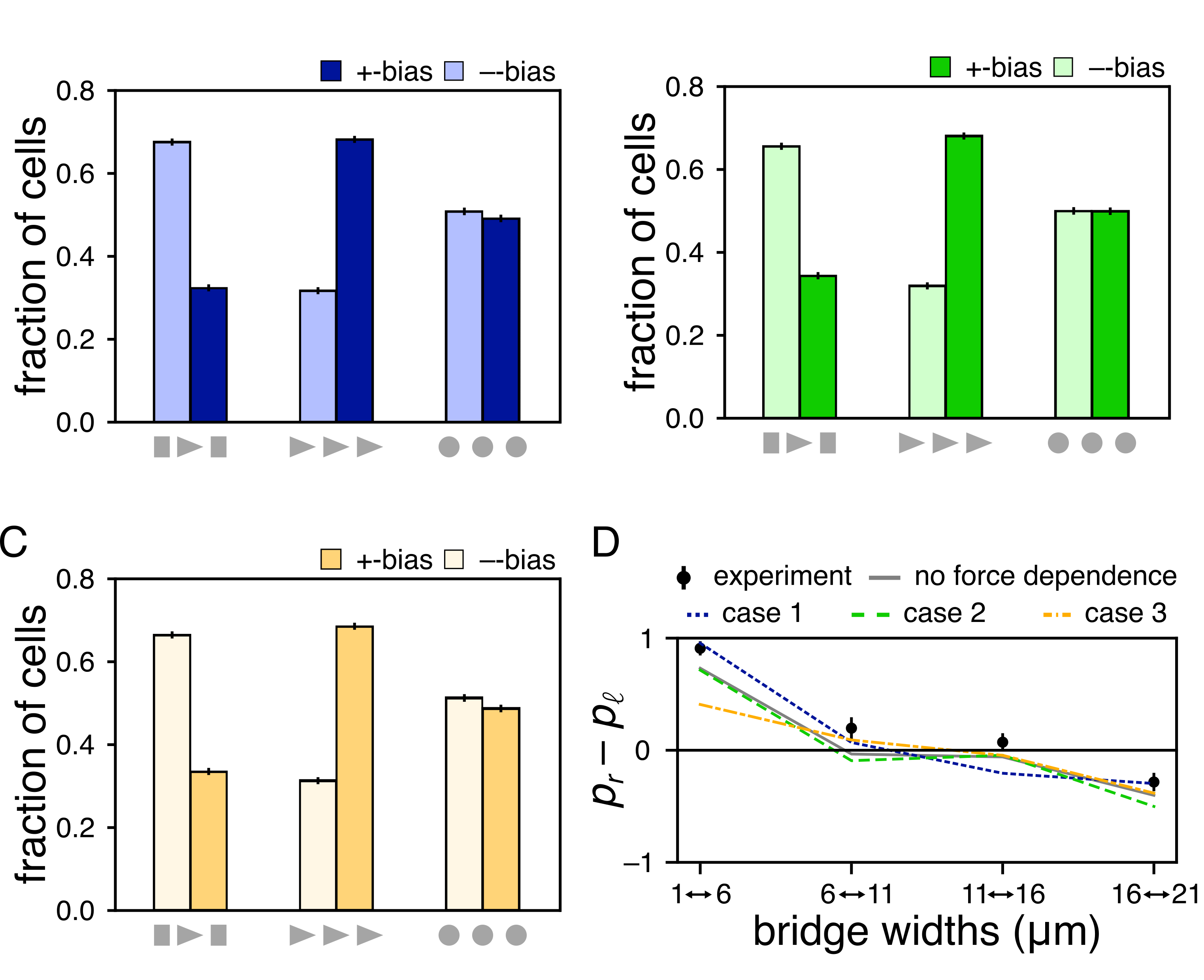}
\caption{\label{fig:figS6} \textbf{Effect of leading order correction terms for actin reorientation and force sensitive binding kinetics on cellular decision making.} \textbf{A-C.} First step migration biases on different asymmetric patterns (see Fig.~2, main text) (A: $k_1\tilde{c}_m+k_2\tilde{c}_\zeta^a > 0$, $\tilde{c}_\zeta^a > 0$; B: $k_1\tilde{c}_m+k_2\tilde{c}_\zeta^a > 0$, $\tilde{c}_\zeta^a < 0$; C: $k_1\tilde{c}_m+k_2\tilde{c}_\zeta^a < 0$, $\tilde{c}_\zeta^a < 0$). \textbf{D.} Migration biases on a chain of
adhesive islands that are connected by adhesive bridges of increasing width (see Fig.~4, main text). Apart from the leading order correction terms, the parameters corresponding to MDA-MB-231 cells on island chains of high adhesiveness were used.}
\end{figure*}

To test if the leading order corrections in Eq.~\eqref{eq:fric_correction}~and~\eqref{eq:fc_corr2} have a qualitative influence on our results in the main text, we performed simulations with three different parameter combinations:
\begin{itemize}
\item case 1: $(k_1 \tilde{c}_m+k_2\tilde{c}_\zeta^a)/\zeta_n(w_\mathrm{cell}) = 3\cdot10^{-4}\text{\textmu m}^{-1}{\rm h}^{-1} > 0$ and $\tilde{c}_\zeta^ak_b/k_\mathrm{off}(0) = 1\cdot10^{-4}\zeta_a(0) > 0$
\item case 2: $(k_1 \tilde{c}_m+k_2\tilde{c}_\zeta^a)/\zeta_n(w_\mathrm{cell}) = 1\cdot10^{-4}\text{\textmu m}^{-1}{\rm h}^{-1} > 0$ and $\tilde{c}_\zeta^ak_b/k_\mathrm{off}(0) = -1\cdot10^{-4}\zeta_a(0) < 0$
\item case 3: $(k_1 \tilde{c}_m+k_2\tilde{c}_\zeta^a)/\zeta_n(w_\mathrm{cell}) = -1\cdot10^{-4}\text{\textmu m}^{-1}{\rm h}^{-1} < 0$ and $\tilde{c}_\zeta^ak_b/k_\mathrm{off}(0) = -3\cdot10^{-4}\zeta_a(0) < 0$
\end{itemize}
While we did observe small quantitative differences between the different parameter combinations, the overall behavior remained qualitatively unchanged (Fig.~\ref{fig:figS6}).

\begin{figure*}[htb]
\includegraphics{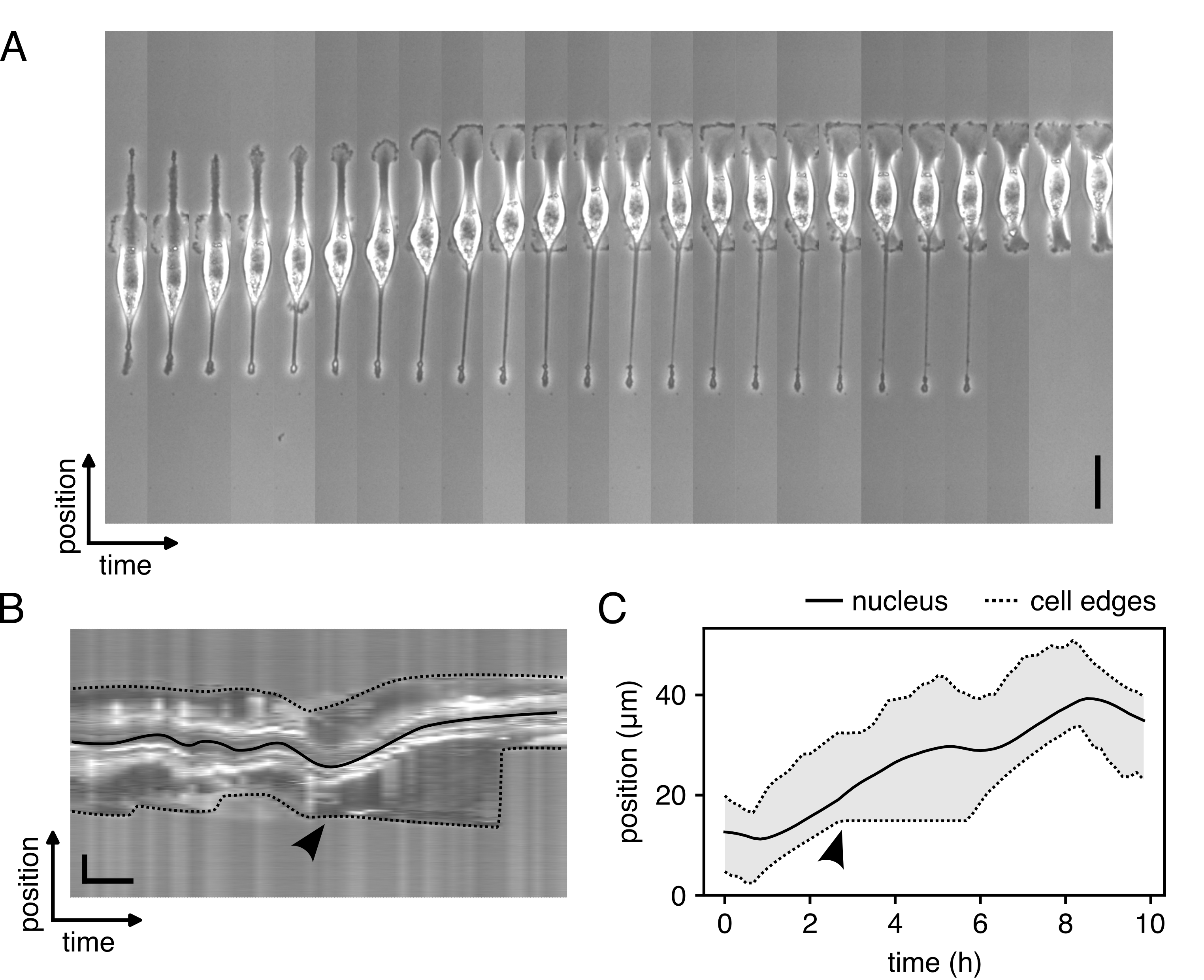}
\caption{\label{fig:figS5} \textbf{Possible effects of myosin unbinding cascades.} \textbf{A.} Time series of a MDA-MB-231 cell with dysfunctional rear contractility (scale bar: 50\textmu m, time interval: 10min). \textbf{B.} The corresponding kymograph (horizontal scale bar: 1h, vertical scale bar: 50\textmu m). \textbf{C.} Simulated kymograph with a model that accounts for load-dependent myosin unbinding cascades. All other parameters are identical to the ones used to model MDA-MB-231 cells on island chains of high adhesiveness. The arrow in B. and C. indicates the time point of rear contraction failure.}
\end{figure*}

For larger forces, the force-dependent unbinding kinetics could lead to unbinding cascades. In the context of adhesion unbinding, this was extensively analyzed in similar models and shown to give rise of so-called stick-slip dynamics~\cite{sens_stickslip_2020, amiri_multistability_2023}. Here, we thus focus on possible myosin unbinding cascades. For this, we build on our intuition obtained from the leading order corrections to our model (Eq.~\eqref{eq:fc_corr2}) from which we saw that the coupling between nucleus and protrusion weakens with increasing load. If the load per myosin filament becomes large enough, this could cause a complete dissociation of myosin filaments. Consequently, the load per remaining myosin filament would spike, resulting in even more filament dissociation. Such a positive feedback would lead to a sudden drop in the contractile force that can be generated by the protrusion. After the tension in the protrusion relaxed, myosin motors could engage again and the contractility of the protrusion would increase again. Here, we account for this in a simplistic way by defining a threshold force per myosin filament $f_\mathrm{thresh}$ that triggers collective myosin dissociation. The load per motor is determined by the retrograde flow velocity and thus the length of the protrusion (see Eq.~(1) and (3), main text). If this velocity exceeds a certain threshold associated with $f_\mathrm{thresh}$, there is a chance that an unbinding cascade occurs, which is modeled by setting the corresponding spring constants temporarily to zero before letting it recover back to its original value. In Fig.~\ref{fig:figS5}, we compare a representative kymograph obtained from simulating this model together with an experimental example showing a similar morphology. Experimentally, such events are however rare, such that we do not expect them to have a significant impact on the population averaged statistics discussed in the main text.

\subsection{The role of boundary effects}
\begin{figure*}[htb]
\includegraphics{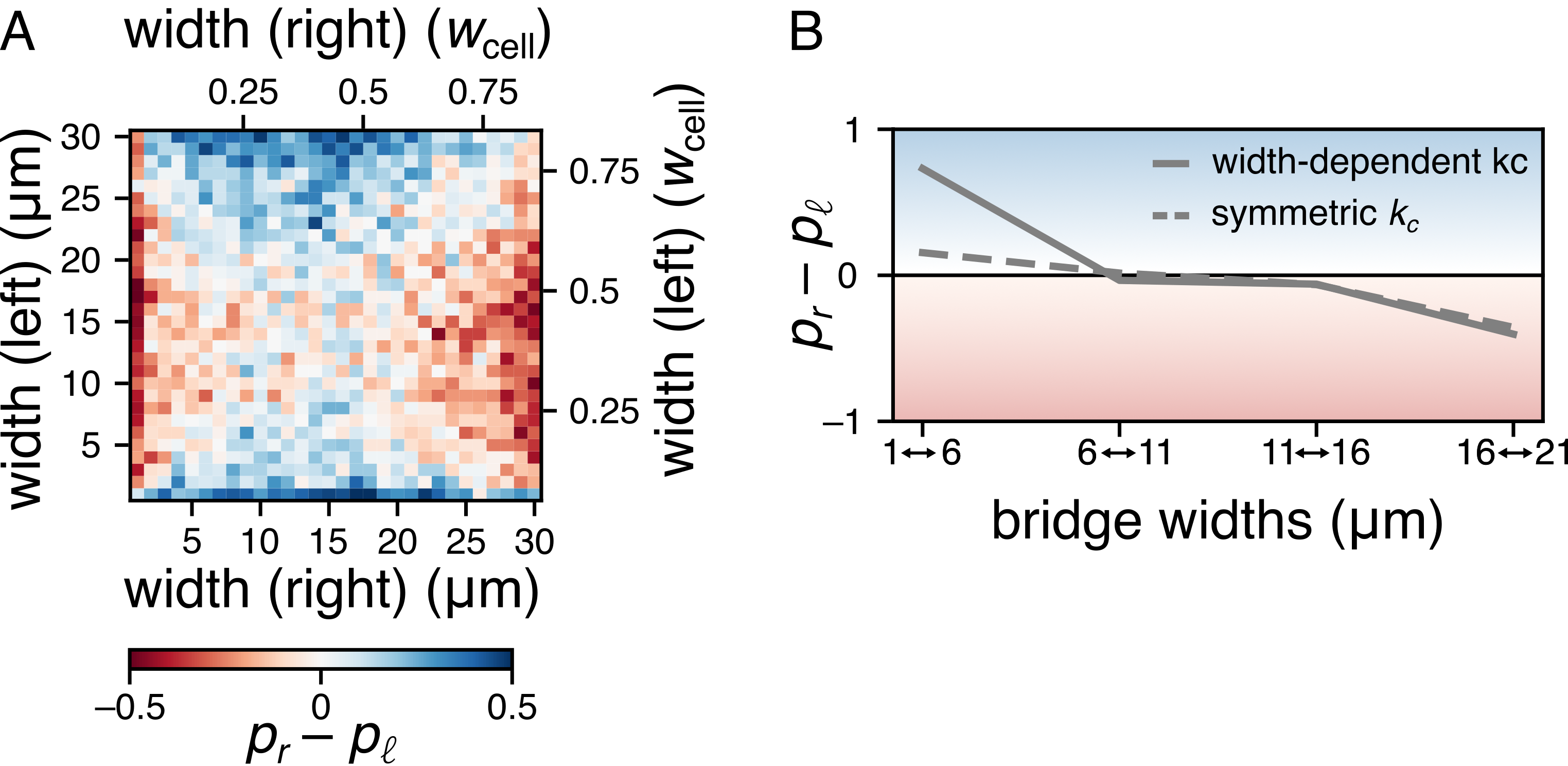}
\caption{\label{fig:figS4} \textbf{Channel width dependence of cellular decision making.} \textbf{A.} Predicted migration bias of a cell seeded on an island with two adjacent, infinitely long channels of different widths. $p_{r/\ell}$ denotes the probability that the cell migrates into the right/left channel. Parameters corresponding to MDA-MB-231 cells on patterns of high adhesiveness were used. \textbf{B.} A model with symmetric, geometry-independent values of $k_c/\zeta_n$ fails to predict the strong bias towards the wider bridge on island 1 on the island chains shown in Fig. 4 in the main text. Solid lines corresponds to the fit of MDA-MB-231 cells on patterns of high adhesiveness, dashed line was obtained with the same parameters except that $k_c = k_c^\mathrm{free}$ was used for all channel widths.}
\end{figure*}

To exclude boundary effects as a possible explanation for the observed switch in preferred migration direction on the chain of islands shown in Fig.~4 in the main text, we simulate our model with symmetric, geometry independent values of $k_c/\zeta_n$ while keeping everything else identical to the simulations shown in Fig.~4B in the main text. We find the width dependence of $k_c/\zeta_n$ to be essential to explain the strong bias towards wider channels on island 1 (Fig.~\ref{fig:figS4}B), indicating is indeed a consequence of the asymmetrically shaped protrusions and not an artifact of the boundaries of the pattern.

\section{Implementation and Model parameters}

\subsection{Effect of channel geometry on the model parameters}
\textbf{Pattern width dependence of $k_c$.} As shown in the main text, the force that a protrusion can apply onto the nucleus scales to leading order linearly with its length (see Eq.~(3)). The effective spring constant $k_c$ that characterizes the coupling between protrusion length and generated force,  however, also depends on the width of the protrusion $w_p$. Using the expression for $k_c$ in the main text Eq.~(3)) together with the expression for the stochastic friction acting on the protrusion (Eq.~\eqref{eq:protrusion_friction}), we find that $k_c\propto w_p$. If the channel is narrow enough, the width of the protrusion is limited to the channel width $w_c$, such that we can write 
\begin{equation}
k_c(w_c) =\begin{cases}
\frac{w_c}{w_\mathrm{sat}}k_c^{\mathrm{free}}, w_c < w_\mathrm{sat}\\
k_c^{\mathrm{free}}, \text{else}
\end{cases}
\end{equation}
where $k_c^{\mathrm{free}}$ denotes the effective spring constant of an unconfined cell and $w_\mathrm{sat}$ is the width at which the pattern width dependence of the effective spring constant saturates. 

There are a number of possible mechanisms leading to such a saturation. First, once the pattern is wider than the unconfined width of the protrusion, further widening of the pattern will not affect the force generation of the protrusion. Second, at some point all available actin and myosin in the protrusion might be used, such that even if there is more space, it is not possible to integrate more actin and myosin into the protrusion. Finally, if the protrusion is significantly larger than the nucleus, only a part of it might be mechanically connected to the nucleus and thus contributing to $f_c$. At this point, it is unclear which of these options is the primary mechanism, such that we treat $w_\mathrm{sat}$ as a free parameter that we fit to experimental data (see section 'Parameter Selection'). 

\textbf{Pattern width dependence of $k_{\ell/r}/\zeta_a$.} 
While the value of $k_c$ and thus also $k_{\ell/r}$ reflects the geometry of the adhesive patch on which the cell body is located, the friction experienced by the protrusion is determined by the adhesive area that is covered by the protrusion. Since both $k_{\ell/r}$ and $\zeta_a$ are proportional to the width of the protrusion (see discussion of $k_c$ above and Eq.~\eqref{eq:actin_friction}), the value of $k_{\ell/r}/\zeta_a$ is independent of the width of the protrusion on homogeneously coated substrates.

\textbf{Pattern width dependence of the nuclear friction.} In the absence of confinement, the friction that the cell experiences, will be determined by the unconfined width of the cell $w_{\rm cell}$, such that $\zeta_n(x_n) = \zeta_n(w_{\rm cell})$. As the cell migrates into a constriction of width $w_c$ that is narrower than the width of the cell, the friction coefficient will gradually decrease until the entire nuclear region is confined to the width $w_c$. To isolate the effect of the confinement, we rewrite the nuclear friction coefficient as $\zeta_n(x_n) = \gamma(x_n)\zeta_n(w_{\rm cell})$, with $\gamma(x_n) = \zeta_n(x_n)/\zeta_n(w_{\rm cell})$.

\begin{figure}[htb]
\includegraphics{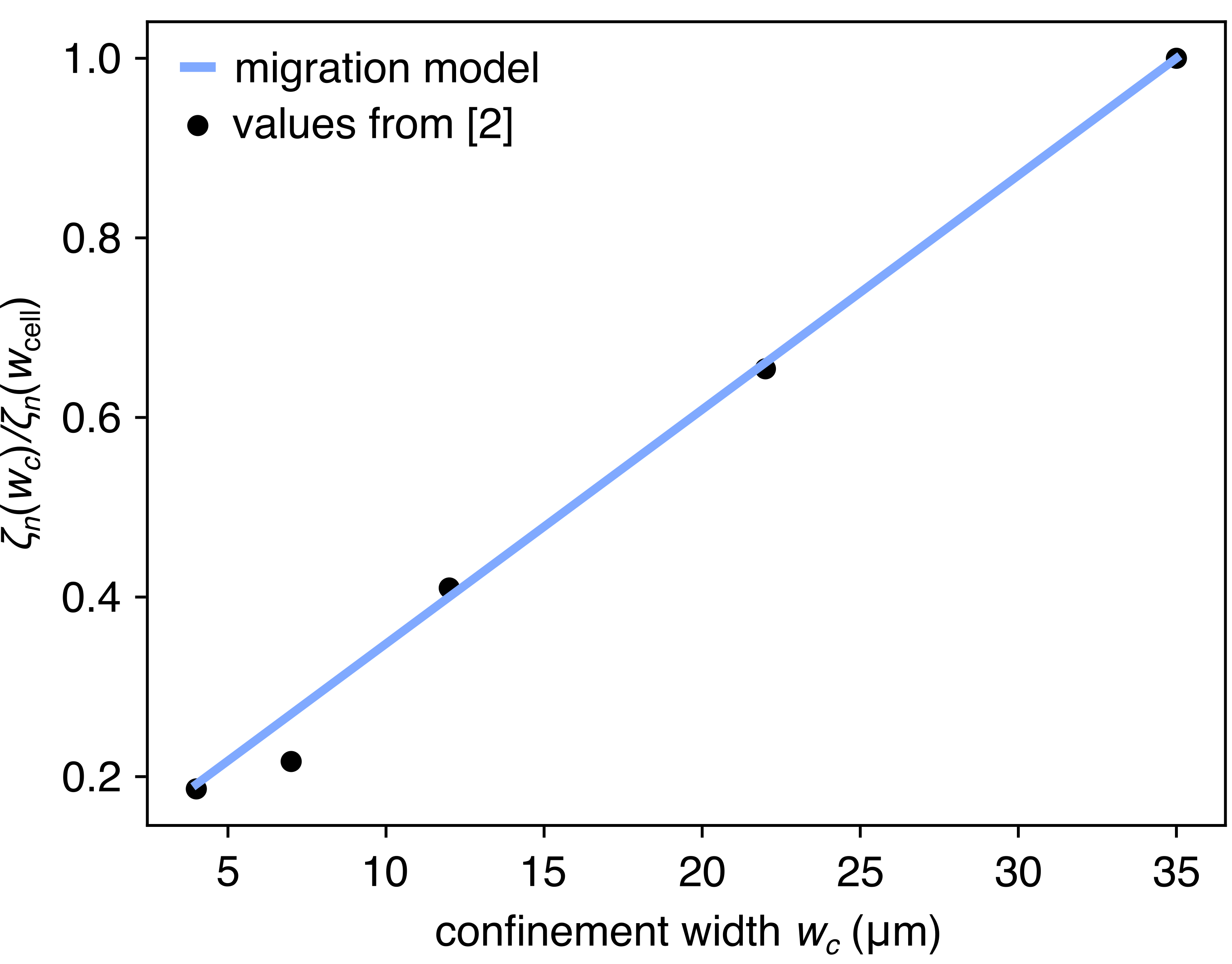}% Here is how to import EPS art
\caption{\label{fig:figS1} \textbf{Nuclear Friction.} Best fit of the minimal value of $\gamma(x_n)$ given by Eq.~\eqref{eq:gamma_min} to the values used in~\cite{bruckner_geometry_2022} that were constrained from experimental data. Fit shown for $\rho_b \ell_n k_b n_0 / (k_\mathrm{off}^0\zeta_0^n)=0.25\ \text{\textmu} {\rm m}^{-1}$ and $w_\mathrm{cell}=35\ \text{\textmu} {\rm m}$.}
\end{figure}

In general, $\gamma(x_n)$ is determined by integrating over the entire nuclear region. Here, we either consider patterns with individual or periodic confinements. In that case, we approximate the complicated spacial dependence of $\gamma(x_n)$ by following our approach from previous work~\cite{bruckner_geometry_2022} as
\begin{equation}\label{eq:gamma}
\gamma(x_n) = \frac{\zeta_n(w_c)}{\zeta_n(w_{\rm cell})} + \frac{1}{2} \left(1 - \frac{\zeta_n(w_c)}{\zeta_n(w_{\rm cell})}\right) \left(1-\cos\left(\frac{2\pi (x_n-x_{\rm center})}{L}\right)\right),
\end{equation}
where $x_{\rm center}$ denotes the center of the confinement and $L$ is the period of the pattern. The bridge width dependence of the minimal value of $\gamma(x_n)$, which is given by 
\begin{equation}\label{eq:gamma_min}
\frac{\zeta_n (w_c)}{\zeta_n(w_{\rm cell})} = \frac{1+\rho_b \ell_n k_b n_0 / (k_\mathrm{off}^0\zeta_0^n) w_c}{1+\rho_b \ell_n k_b n_0 / (k_\mathrm{off}^0\zeta_0^n) w_\mathrm{cell}},
\end{equation}
is in excellent agreement with the values found in~\cite{bruckner_geometry_2022} by fitting the experimentally observed nuclear velocities for varying bridge widths (see Fig.~\ref{fig:figS1}). 

\subsection{Effect of the adhesion density on the model}
The density of adhesive molecules on the substrate $\rho_\mathrm{ad}$ will determine the number of adhesive bonds formed. In the absence of saturation effects, we expect that the bond density $\rho_b \propto \rho_\mathrm{ad}$. From Eq.~(3) in the main text, we get that 
\begin{equation}\label{eq:kc}
k_c = \frac{4\tau_1\tilde{N}_F\rho_b\rho_m w_pf_s}{\left(\zeta_a - \tilde{N}_F\rho_b\rho_mw_p f_s\right)h_p} w_p.
\end{equation}
Using that $\zeta_a = \zeta_1^a(0) = \rho_b \ell_r k_b n_0 w_p / k_\mathrm{off}^0$ (see Eq.~\eqref{eq:protrusion_friction}), we find that $\rho_b$ cancels out in Eq.~\eqref{eq:kc} and $k_c$ is thus independent of the adhesiveness of the pattern. Similarly, also the effective spring constants $k_{\ell/r} = k_c+4\tau_1w_p/h_p$ are independent of $\rho_b$. 

This leaves us with the two friction coefficients $\zeta_a$ and $\zeta_n$ as $\rho_b$ sensitive parameters. As discussed in the Section 'Internal and stochastic friction', the contribution of the viscous drag to the total drag coefficient of the actin filaments can be neglected. As a consequence, $\zeta_a \propto \rho_b$ and the rescaled spring constants $k_{\ell/r}/\zeta_a \propto \rho_b^{-1}$. In contrast, for the nuclear friction the viscous drag has to be taken into account. Using Eq.~\eqref{eq:gamma_min}, we can write the contribution of the viscous drag to the nuclear friction coefficient $\zeta_0^n$ as
\begin{equation}
\zeta_0^n = \frac{\zeta_0^n}{\zeta_n(w_\mathrm{cell})}\zeta_n(w_\mathrm{cell}) = (1 + \rho_b \ell_n k_b n_0 / (k_\mathrm{off}^0\zeta_0^n) w_\mathrm{cell})^{-1}\zeta_n(w_\mathrm{cell})
\end{equation}
and thus
\begin{equation}
\zeta_1^n(w_\mathrm{cell}) = \left(1 - (1 + \rho_b \ell_n k_b n_0 / (k_\mathrm{off}^0\zeta_0^n) w_\mathrm{cell})^{-1}\right)\zeta_n(w_\mathrm{cell}).
\end{equation}
Using that $\zeta_1^n=\rho_b \ell_n k_b n_0 w_n / k_\mathrm{off}^0$ we get
\begin{equation}\label{eq:rho_b_dep_zeta_n1}
\zeta_n(w_\mathrm{cell}) = \frac{\rho_b \ell_n k_b n_0 w_\mathrm{cell} / (k_\mathrm{off}^0\zeta_0^n)}{1 - (1 + \rho_b \ell_n k_b n_0 w_\mathrm{cell} / (k_\mathrm{off}^0\zeta_0^n))^{-1}}\zeta_0^n.
\end{equation}
Based on the fit shown in Fig.~\ref{fig:figS1}, we find that $\rho_b \ell_n k_b n_0 w_\mathrm{cell} / (k_\mathrm{off}^0\zeta_0^n) = 8.75\rho_b(\rho_\mathrm{ad})/\rho_b(\rho_\mathrm{ad}^\mathrm{ref})$, where the reference concentration $\rho_\mathrm{ad}^\mathrm{ref}$ denotes the fibronectin concentration for the experiments with high adhesiveness (Fig.~4B, main text). We can thus rewrite Eq.~\eqref{eq:rho_b_dep_zeta_n1} as
\begin{equation}\label{eq:rho_b_dep_zeta_n2}
\zeta_n(w_\mathrm{cell}, \rho_\mathrm{ad}) = \frac{8.75\rho_b(\rho_\mathrm{ad})/\rho_b(\rho_\mathrm{ad}^\mathrm{ref})}{1 - (1 + 8.75\rho_b(\rho_\mathrm{ad})/\rho_b(\rho_\mathrm{ad}^\mathrm{ref}))^{-1}}\zeta_0^n = \frac{\rho_b(\rho_\mathrm{ad})}{\rho_b(\rho_\mathrm{ad}^\mathrm{ref})}\frac{1 - (1 + 8.75)^{-1}}{1 - (1 + 8.75\rho_b(\rho_\mathrm{ad})/\rho_b(\rho_\mathrm{ad}^\mathrm{ref}))^{-1}}\zeta_n(w_\mathrm{cell}, \rho_\mathrm{ad}^\mathrm{ref}).
\end{equation}

Based on this, we predict the values of the rescaled spring constant for the experiments with reduced adhesiveness~($\rho_\mathrm{ad} = 0.5 \rho_\mathrm{ad}^\mathrm{ref}$) as given in Table~\ref{tab:table1}.

\subsection{Effect of ratchet geometry on the model}
\textbf{Geometry dependence of $k_c$.} As discussed in the main text, ratchet-like geometries have two opposing effects on protrusion growth. First, on the wider side ($-$-direction), the cell can form wider/more protrusions. As discussed in the previous section, $k_c$ scales with the pattern width as $k_c\propto w_p$.

As a consequence, on ratchet patterns, we expect $k_c$ to be larger in the direction of the blunt side ($-$-direction) of the pattern compared to the pointed side ($+$-direction). In the context of this paper, we always orient ratchet-shaped patterns such that the pointed side of the pattern points in the right direction. Consistent with previous work~\cite{caballero_protrusion_2014}, we thus account for the ratchet-shaped starting pattern by choosing $k_c(x_\ell) > k_c(x_r)$ in both the single ratchet as well as the periodic ratchet case (see Table~\ref{tab:table1}). In the control case of periodic spherical patterns, the protrusion on both sides of the cell can be equally wide, such that we choose $k_c(x_\ell) = k_c(x_r)$.

\textbf{Geometry dependence of $k_{\ell/r}/\zeta_a$.} As discussed above, $k_{\ell/r}/\zeta_a$ is independent of the width of the protrusion on homogeneously coated substrates. The only remaining dependence on the geometry of the adhesive pattern is then the density of adhesive bonds $\rho_b$ in Eq.~\eqref{eq:actin_friction}.

In both the experiments in \cite{caballero_protrusion_2014} and our simulations, the cells are seeded on the central patch and then migrate by extending protrusions to the neighboring patches before eventually pulling the nucleus over to one of the neighboring patterns. By doing so, parts of the protrusions overlap with non adhesive surface regions. The fraction of adhesive and non-adhesive surface area covered by the protrusions then determines the average adhesive bond density $\rho_b$. In the case of the single ratchet and the symmetric circular pattern, the adhesive region below the front of the protrusions are identical on both sides of the cell, since the neighboring patterns are identical in both directions. As a consequence $k_{\ell}/\zeta_a=k_{r}/\zeta_a$. In contrast, in the case of periodic ratchets, the protrusion pointing in the $+$-direction overlaps with the pointed end of the neighboring pattern, which results in a lower average adhesive bond density compared to the adhesion in the $-$-direction (see Fig.~2A). To account for this, we choose $\zeta_a(x_\ell) < \zeta_a(x_r)$.

To analyze how the geometry affects the migration behavior, we increase both $k_\ell/\zeta_a$ and $k_c(x_\ell)/\zeta_n$ while keeping $k_r/\zeta_a$ and $k_c(x_r)/\zeta_n$ fixed and calculate the resulting average migration bias (see Fig.~\ref{fig:figS2}A). We find that increasing $k_c(x_\ell)/\zeta_n$ leads to a migration bias in the $-$-direction, while  additionally increasing $k_\ell/\zeta_a$ counteracts this effect and if chosen high enough results in a bias in the $+$-direction. The parameter values in Table~\ref{tab:table1} were chosen such that the biases predicted by our model on the different patterns were in qualitative agreement with experimental data in~\cite{caballero_protrusion_2014, lo_vecchio_collective_2020} (Fig. 2B and C, main text).

\textbf{Geometry dependence of $\zeta_n$.} The stochastic friction experienced by the nucleus also also depend on the width of the pattern. In particular, the friction in the non-adhesive region between two patterns should be minimal. We incorporate this by gradually decreasing the nuclear friction coefficient according to Eq.~\eqref{eq:gamma} from its maximal value in the center of the adhesive patch to its minimum between two patches. In addition, on ratchet geometries, we would in principle expect that the friction varies differently towards both sides of the pattern. We find however that asymmetric friction profiles do not affect our results qualitatively such that we neglect this effect in the interest of simplicity of the model and choose a symmetric profile given by Eq.~\eqref{eq:gamma}.

\textbf{Actin alignment on ratchet geometries.}
In the case of pattern boundaries that are parallel to the protrusion-nucleus axis, we expect that the boundaries lead to an increased alignment of actin in the direction of migration with decreasing pattern width. On ratchet geometries however, the boundaries of the patterns are not aligned with the direction of protrusion growth. Additionally, the pattern boundaries on both sides of the protrusion point in opposing directions inhibiting long range alignment of actin filaments in the protrusion. We thus expect that this effect has much less impact on the polarization dynamics than the protrusion width and the density of adhesive bonds. This is also consistent with previous analysis of the migration behavior on these geometries~\cite{caballero_protrusion_2014}. We thus choose identical $S_{\ell/r}$ on both sides of the cell and absorb the its value into the polymerization rate $r_p$, such that $s=0$ and thus $S_\ell=S_r=1$ (see Table~\ref{tab:table1}).

\textbf{Ratchet spacing.} One experimental parameter that can be tuned to check the validity of our model is the spacing between neighboring patches. This affects how much the protrusion overlaps with the adhesive pattern. Hence, the friction of the protrusion $\zeta_a$ decreases with increasing pattern spacing. 

The adhesive area below the protrusion $A_\mathrm{ad}$ is different depending on what side of the ratchet-shaped patch the protrusion encounters. Following the orientation shown in Fig.~2A in the main text, the right protrusion encounters the blunt end of the pattern while the left protrusion overlaps with the pointed end. At zero pattern spacing, we then write the adhesive area in terms of protrusion dimensions as
\begin{equation}
A_\mathrm{ad} = g_{\ell/r}A_p \approx g_{\ell/r}w_pL_p,
\end{equation}
where the geometric factor $0 < g_{\ell/r} \leq 1$ accounts for the shape of the pattern and $A_p$ denotes the area of the protrusion. Since the blunt end of the pattern is typically wider than the protrusion such that the entire right protrusion is on an adhesive surface, we choose $g_r=1$ and $g_\ell < 1$. For finite pattern spacing $\Delta x_\mathrm{pattern} > 0$, the adhesive area below the protrusion is reduced since the protrusion needs to span over the non-adhesive gap between neighboring patches, such that we get
\begin{equation}
A_\mathrm{ad} = g_{\ell/r}(L_p-\Delta x_\mathrm{pattern})w_p.
\end{equation}
The average density of adhesive molecules below the protrusions is then given by
\begin{equation}\label{eq:spacing}
\rho_\mathrm{ad} = \rho_\mathrm{ad}^\mathrm{hom}\frac{A_\mathrm{ad}}{A_p} = \rho_\mathrm{ad}^\mathrm{hom} g_{\ell/r}(1-L_p^{-1}\Delta x_\mathrm{pattern}),
\end{equation}
where $\rho_\mathrm{ad}^\mathrm{hom}$ denotes the average adhesion density on a homogeneously coated surface. It was previously shown that the number of adhesion bonds that cells can form saturates at high numbers of adhesive molecules and that this was essential to explain the migration behavior of NIH3T3 cells on ratchet geometries~\cite{lo_vecchio_collective_2020}. We account for this by relating the bond density $\rho_b$ to the density of adhesive molecules below the protrusion $\rho_\mathrm{ad}$ through
\begin{equation}\label{eq:saturation}
\rho_\mathrm{b} = \rho_\mathrm{b}^\mathrm{hom}\tanh\left(\frac{\rho_\mathrm{ad}}{\rho_\mathrm{sat}}\right),
\end{equation}
where $\rho_\mathrm{sat}$ determines how fast the bond density saturates and we assumed that the density of the pattern is high enough such that the bond density on homogeneously coated surfaces $\rho_b^\mathrm{hom}$ is equal to the saturation density, which was shown to be the case in~\cite{lo_vecchio_collective_2020}.

Using Eqs.~\eqref{eq:spacing} and \eqref{eq:saturation} together with Eq.~\eqref{eq:protrusion_friction}, we can express the rescaled spring constants $k_{\ell/r}/\zeta_a$ in terms of the pattern spacing $\Delta x_\mathrm{pattern}$ as 
\begin{equation}\label{eq:kp_spacing}
\frac{k_{\ell/r}}{\zeta_a} = \frac{k_{\ell/r}}{\zeta_a^\mathrm{hom}}\frac{1}{\tanh\left(\frac{g_{\ell/r}\rho_\mathrm{ad}^\mathrm{hom}}{\rho_\mathrm{sat}}(1-L_p^{-1}\Delta x_\mathrm{pattern}\right)},
\end{equation}
where $\zeta_a^\mathrm{hom}$ is the effective friction coefficient of the protrusion on a homogeneously coated substrate.

\begin{figure}[htb]
\includegraphics{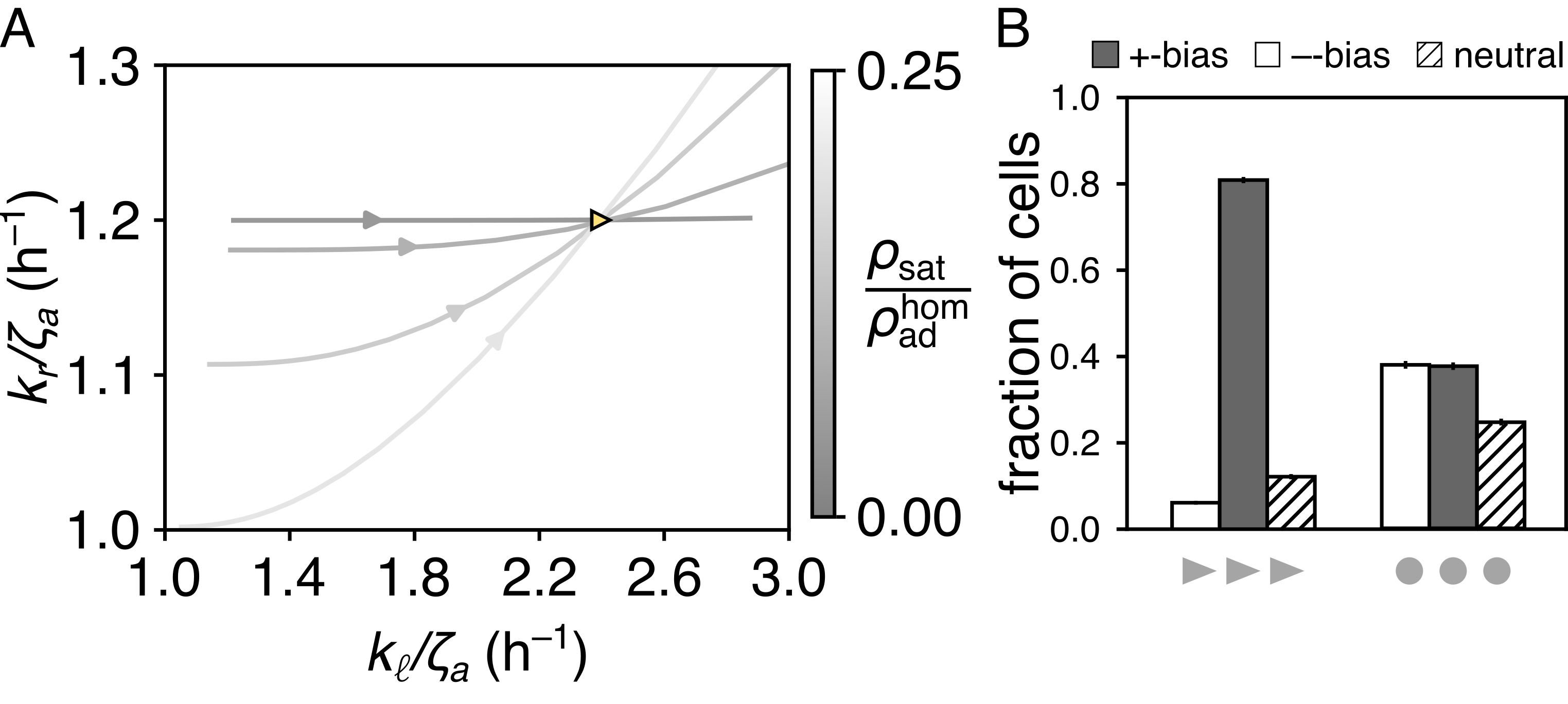}% Here is how to import EPS art
\caption{\label{fig:figS7} \textbf{Long term dynamics of the model on ratchet geometries.} \textbf{A.} Dependence of the rescaled spring constants on pattern spacing for different values of the saturation parameter $\rho_\mathrm{sat}/\rho_\mathrm{ad}^\mathrm{hom}$ (0.05, 0.10, 0.15, 0.20). With increasing pattern spacing both $k_\ell/\zeta_a$ and $k_r/\zeta_a$ increase, but $k_\ell/\zeta_a$ is much more sensitive to the pattern spacing due to the asymmetric shape of the adhesive patches. The parameter combination chosen to fit migration bias at a pattern spacing of 20.5 \textmu m in the main text is indicated by the yellow triangle. \textbf{B.} Predicted biases after 48h on ratchet and circular patterns. On ratchets, the model predicts a large fraction of cells being biased in the $+$-direction.}
\end{figure}

For the experiments in~\cite{caballero_protrusion_2014}, a pattern spacing of $\Delta x_\mathrm{pattern} = 20.5$\textmu m was used and $L_p$ was found to be approximately 27\textmu m. By imposing that at a pattern spacing of 20.5\textmu m the rescaled spring constants match the values that were chosen to fit the data in~\cite{caballero_protrusion_2014, lo_vecchio_collective_2020} at that width ($k_r/\zeta_a=1.2$ ${\rm h}^{-1}$ and $k_\ell/\zeta_a=2.3$ ${\rm h}^{-1}$, see Fig.~2B and C main text), we constrain the parameters $k_{\ell/r}/\zeta_a$ and $g_\ell$ (see Fig.~\ref{fig:figS7}). This leaves us with a single free parameter $\rho_\mathrm{sat}/\rho_\mathrm{ad}^\mathrm{hom}$ that determines the onset of saturation. For a value of $\rho_\mathrm{sat}/\rho_\mathrm{ad}^\mathrm{hom} = 0.05$ ($g_\ell=0.12$, $k_\mathrm{\ell/r}/\zeta_a^\mathrm{hom} = 1.2$ ${\rm h}^{-1}$), we find good agreement with the experimental data in~\cite{lo_vecchio_collective_2020}  (Fig. 2C, main text).

\subsection{Cell-to-cell variability}
For migration on ratchet-like geometries, we chose model parameters that could qualitatively reproduce the first step biases (Fig. 2B) and quantitatively fit the $\bar{p}$ values for varying pattern spacing (Fig. 2C). Based on this, we predicted the long term migration biases on periodic patterns (Fig.~\ref{fig:figS7}B). While our model qualitatively predicts the correct bias, we observe a over-representation of $+$-biased cells compared to~\cite{caballero_protrusion_2014}. A part of this can be attributed to stationary cells in the experiment ($\sim$5\% of the cell population). These cells were not included in the first step migration biases or the $\bar{p}$ value and were thus not accounted for in our parameter choice. 

The quantitative agreement of the $\bar{p}$ values together with an over-representation of $+$-biased cells indicates that cell-to-cell variability plays a significant role in this system: A lower fraction of $+$-biased cells with the same overall $\bar{p}$ value could be achieved by having fewer, but more biased $+$-biased and consequently more $-$-biased cells in the population. A better representation of the full cell population could thus require to simulate our model with a range of different parameter values accounting for differences in the internal states of cells. In previous work from our groups, we showed that such an approach can improve model predictions on the population level for cells migrating on dumbbell-shaped micro-patterns~\cite{bruckner_disentangling_2020}.

\subsection{Numerical Implementation}
To numerically solve the closed set of Eqs.~(4),~(5)~and~(8), we implemented them in Python 3.7 and integrated them with a forward Euler scheme. We simulated a population of cells by repeated simulations on the pattern with random initialization. For the simulations on dumbbells and island chains, we simulated 300 trajectories of 50h, where we recorded the degrees of freedom every 10 min to ensure comparability to the experimental data. For determining the first step migration bias on periodic patterns (Fig. 2B), we simulated 3000 cells with $\Delta x_\mathrm{pattern}$ = 20.5\textmu m and terminated the simulations after cells moved their nucleus to a neighboring pattern. For the 48h migration biases (Fig.~\ref{fig:figS7}B) we simulated 3000 cells with $\Delta x_\mathrm{pattern}$ = 20.5\textmu m and terminated the simulations after 48h and classified cells according to their final position relative to their starting position. Cells were only considered to show a biased migration if the final location of their nucleus was not on their starting pattern. For the calculation of the $\bar{p}$-values for varying pattern spacing (Fig. 2C), we simulated 300 trajectories of 50h with varying $\Delta x_\mathrm{pattern}$ and counted a step whenever the nucleus moved onto a neighboring patch.

For the non-periodic patterns (dumbbells and island chain), cells cannot grow effective protrusion beyond the outer boundaries of the pattern. We implement this in our model by setting the polymerization rate to zero whenever a protrusion coordinate reaches the edge of the pattern.

\subsection{Parameter selection}
To determine the parameter values used in the simulations, we fixed as many parameters as possible either through the experimental geometry ($L$, $\Delta x_\mathrm{pattern}$) or by comparison to a model of similar structure that we previously constrained from experimental data (spring constants, friction coefficients, parameters that determine $\alpha$, $\beta$ and $\sigma$) \cite{bruckner_geometry_2022}. The polymerization rate was chosen such that the cellular dimensions approximately agree with the experiments. The parameter $s$ - which characterizes the coupling between pattern width and actin alignment - and the bias strength $\delta$ were chosen to match experimental statistics. 

Depending on the geometry and considered cell line, we then deviated from these values to account for asymmetries in the adhesive patterns (see discussion above), differences between cell lines or in the case of the island chains, to ensure a better agreement between experiment and simulations. The fact that we find slightly different parameters to work for the island chains compared to the dumbbells could be due to slight differences in the experimental design between this work and~\cite{bruckner_stochastic_2019}. The used values for the different model parameters are summarized in Table~\ref{tab:table1}.

\begin{table*}[hbt]
\centering
\begin{tabular}{|l||c|c|c|c|c|c|c|}
 \hline
 geometry & single  & periodic  & periodic  & dumbbell & island chain, MDA & island chain, HT & island chain, MDA\\
  & triangle & triangle & circles & & (high adhesiveness) & (high adhesiveness) & (low adhesiveness) \\
  \hline\hline
  $k_\ell/\zeta_a$ (${\rm h}^{-1}$) & 1.2 & see Eq.~\eqref{eq:kp_spacing} & 1.2 & 1.2 & 1.4 & 1.2 & 2.8\\
  $k_r/\zeta_a$ (${\rm h}^{-1}$) & 1.2 & see Eq.~\eqref{eq:kp_spacing} & 1.2 & 1.2 & 1.4 & 1.2 & 2.8\\
  $k_c(x_\ell)/ \zeta_n(w_{\rm cell})$ (${\rm h}^{-1}$) & 0.8 & 0.8 & 0.6 & 0.6 & 0.8 min$\left(\frac{w_p(x_\ell)}{14}, 1\right)$ & 0.6 min$\left(\frac{w_p(x_\ell)}{19}, 1\right)$ & 1.45 min$\left(\frac{w_p(x_\ell)}{14}, 1\right)$ \\
  $k_c(x_r)/ \zeta_n(w_{\rm cell})$ (${\rm h}^{-1}$)& 0.6 & 0.6 & 0.6 & 0.8 & 0.8 min$\left(\frac{w_p(x_\ell)}{14}, 1\right)$ & 0.6 min$\left(\frac{w_p(x_\ell)}{19}, 1\right)$ & 1.45 min$\left(\frac{w_p(x_\ell)}{14}, 1\right)$ \\
  $\zeta_n(w_c)/\zeta_n(w_{\rm cell})$ & 0.2 & 0.2 & 0.2 & $\frac{1+0.25w_c}{1+0.25\cdot 35}$ & $\frac{1+0.25w_c}{1+0.25\cdot 35}$ & $\frac{1+0.25w_c}{1+0.25\cdot 35}$ & $\frac{1+0.125w_c}{1+0.125\cdot 35}$ \\
  $L$ (\textmu m) & 127 & 106.5 + $\Delta x_\mathrm{pattern}$ & 127 & 52.5 & 85 & 85 & 85\\ 
  $\ell_a r_p(c_\mathrm{0})$ (\textmu ${\rm m h}^{-1}$) & 40 & 40 & 40 & 5 & 20 & 20 & 20 \\
  $D - \frac{8n_c c_0 \ell_a r_1}{L_c}$ (${\rm h}^{-1}$) & 1.5 & 1.5 & 1.5 & --22.9 & --21.5 & --12 &  --21.5\\
  $\frac{4n_c c_0\ell_a r_1s}{L_c}$ ($\text{\textmu} {\rm m}^{-2}{\rm h}^{-1}$) & 0 & 0 & 0 & 0.0135 & 0.014 & 0.0076 & 0.014 \\
  $s$ ($\text{\textmu} {\rm m}^{-2}$) & 0 & 0 & 0 & 0.0003 & 0.0003 & 0.0003 & 0.0003 \\
  $\beta$ ($\text{\textmu} {\rm m}^{-2}{\rm h}$) & 0 & 0 & 0 & 0.001 & 0.001 & 0.001 & 0.001 \\
  $\delta$ (${\rm h}^{-2}$) & 0 & 0 & 0 & 150 & 150 & 65 & 150 \\
  $\sigma$ ($\text{\textmu} {\rm mh}^{-3/2}$) & 100 & 100 & 100 & 100 & 100 & 100 & 100\\ 
  $l_ar_1P_0^{-1}$ & 1 & 1 & 1 & 1 & 1 & 1 & 1\\ 
\hline
\end{tabular}
\caption{\label{tab:table1} Model parameters used to describe the different experiments if not explicitly stated otherwise.}
\end{table*}

\subsection{Reduced phase space analysis}
Even though the parameter space of our model is high dimensional and not every biological parameter can be individually constrained from our data, the effect of most experimentally relevant parameters can be grouped into two qualitatively distinct aspect of cellular migration behavior: effective contractility and polarization. This allows us to qualitatively discuss the effects of a number of key biological parameters such as the number of actin and myosin, the adhesiveness and the concentration of the polarity cue on the migration behavior in a reduced two dimensional phase space.

The effective contractility of cells is characterized in our model by four rescaled spring constants (see Table~\ref{tab:table1}) that all consist of an effective spring constant that accounts for actomyosin contractility and membrane tension as well as a friction coefficient that accounts for the stochastic (un-)binding kinetics of adhesion bonds and viscous drag. The value of these rescaled spring constants is determined by parameters such as the number of actin filaments $N_F^{n/p}$ or the myosin concentration $\rho_m$ in the cytosol as well as the adhesion density $\rho_b$. These parameters are likely to vary between different cell lines and experimental conditions due to differences in gene expression levels or the properties of the environment. While increased actin and myosin concentrations will generally lead to an increased contractility, increased adhesiveness has the opposite effect on the rescaled spring constants and thus decrease the effective contractility. The polarization dynamics of cells arise from the balance between diffusion and advection of the polarity cue in our model. The key biological parameters that control these processes are the diffusion coefficient in the cytosol $\tilde{D}$, the overall polarity cue concentration $c_0$ and the fraction of advected polarity cue $n_c$. Generally, an increased polarity concentration and advected fraction will lead to an increase in polarity while larger diffusivity will decrease polarity. For a more detailed analysis of the sensitivity of cell migration behavior on parameter values, we will consider the cases of migration on directed adhesive patterns and island chains separately.

\begin{figure*}[htb]
\includegraphics{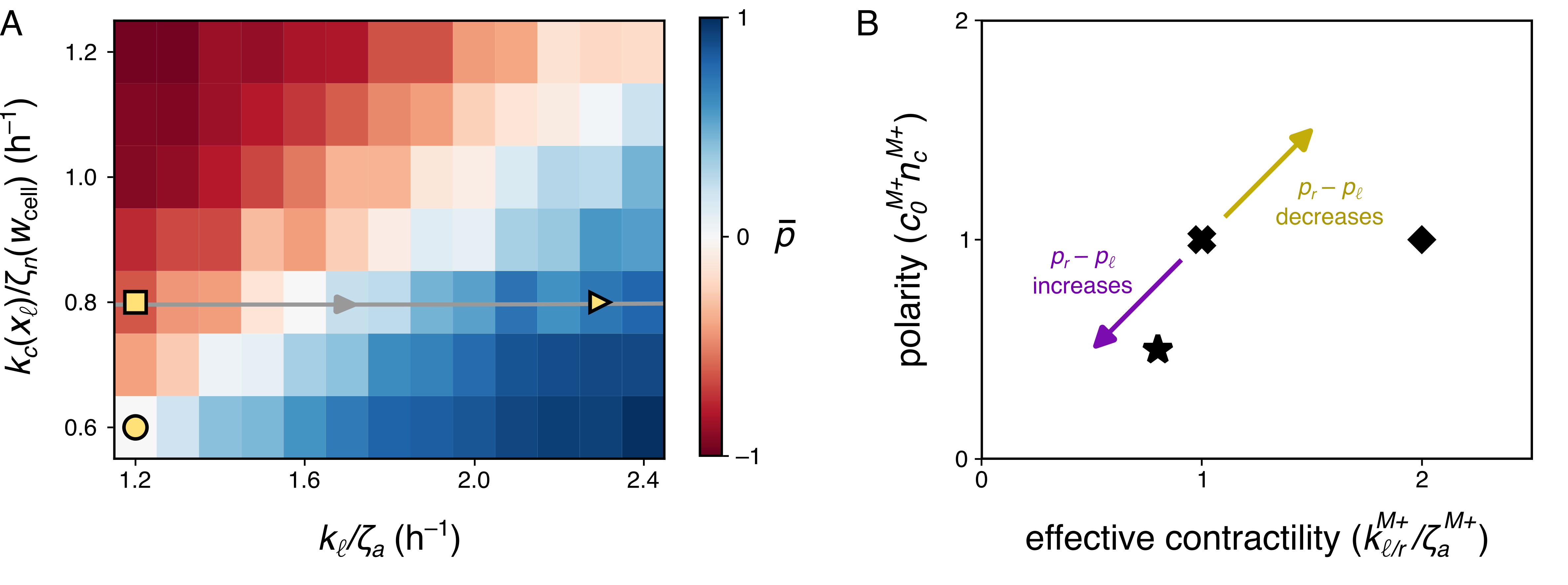}
\caption{\label{fig:figS2} \textbf{Reduced phase space analysis.} \textbf{A.} Parameter sweeps of the model with asymmetric protrusions on ratchet-like patterns. The pattern geometry affects the rescaled effective spring constants $k_c(x_{\ell/r})/\zeta_n(w_\mathrm{cell})$ and $k_{\ell/r}/\zeta_a(x_{\ell/r})$ and thus the overall migration bias $\bar{p}$ (see main text). To explore the effect of the asymmetry in rescaled spring constants, we keep $k_c(x_{r})/\zeta_n(w_\mathrm{cell})$ and $k_{r}/\zeta_a(x_{r})$ fixed while varying the other parameters. The experimental conditions shown in Fig.~2B in the main text are indicated by yellow markers (periodic circles pattern: circle, single triangle pattern: square, periodic triangle pattern: triangle). The gray arrow indicates the effect of pattern spacing on the rescaled spring constants for periodic triangular patterns. \textbf{B.} Reduced phase space of our model on island chains. The migration behavior of cells on island chains can be qualitatively characterized in the contractility-polarity space. The estimated position of the different island chain experiments relative to the case of MDA-MB-231 cells on highly adhesive patterns (M+) are indicated by the black markers (cross: MDA-MB-231 cells on highly adhesive patterns, diamond: MDA-MB-231 cells on less adhesive patterns, star: HT-1080 cells on highly adhesive patterns).}
\end{figure*}

\textbf{Asymmetric contractility on directed adhesive patterns.} As discussed in the main text, asymmetries between the rescaled spring constants associated with both sides of the nucleus can give rise to biased migration. Migration on 'ratchet'-like adhesive patterns is particularly suited to discuss the effect that asymmetric contractility has on cell migration, since complex polarization dynamics are not required to explain the observed migration behavior. The adhesive patterns in~\cite{caballero_protrusion_2014, lo_vecchio_collective_2020} were designed and oriented such that effective contractility on the left side of the nucleus is increased compared to the right side (Fig. 2, main text). Depending on which of the two rescaled spring constants $k_c(x_\ell)/\zeta_n(w_\mathrm{cell})$ or $k_\ell/\zeta_a$ is increased, this can result in qualitatively different migration behavior (Fig.~\ref{fig:figS2}A): Increasing only the effective spring constant $k_c(x_\ell)/\zeta_n(w_\mathrm{cell})$ experienced by the nucleus as realized through forcing the formation of protrusion of different widths $w_p$ on single triangle patterns (see Section 'Effect of ratchet geometry on the mode') leads to a net force towards the left ($-$) side of the cell resulting in biased migration towards the left (red region in Fig.~\ref{fig:figS2}A). In contrast, selectively increasing of the effective spring constant $k_\ell/\zeta_a$ experienced by the front of the protrusion through an asymmetry in the adhesion density $\rho_b$ leads to a shorter protrusion on the left side of the nucleus resulting in a net force in the opposite direction and thus biased migration towards the right (blue region in Fig.~\ref{fig:figS2}A). If both effective spring constants associated with the left side of the cell are increased such as in the case of migration on periodic triangular patterns the balance between the two dictates the overall bias. This can be controlled experimentally by varying the pattern spacing~\cite{lo_vecchio_collective_2020} resulting in a stronger increase of $k_\ell/\zeta_a$ with increasing pattern spacing compared to $k_c(x_\ell)/\zeta_n(w_\mathrm{cell})$ (Fig.~\ref{fig:figS7}~and~\ref{fig:figS2}A).

\textbf{Effective contractility and polarization on island chains.}
For migration on island chains, the effect of asymmetric contractility competes with confinement-induced polarization dynamics, resulting in different migration biases depending on the width of the two neighboring channels (Fig.~\ref{fig:figS4}A). This allows for an analysis of the effect of overall contractility and the polarization strength on the balance between these two counteracting effects. Of the three biological parameters discussed above that control the effective contractility (actin and myosin concentration and density of adhesive bonds), the adhesion density is the easiest to control experimentally by varying the adhesiveness of the substrate. Both experimentally and through simulations, we find that while an increase of the effective contractility increases the absolute difference between the rescaled spring constants on both sides of the cell, it also increases the effect of the polarization on the migration dynamics, since differences in protrusion length lead to higher force differences. For MDA-MB-231 cells, this leads to an increased bias of the migration towards the narrower channel on substrates of reduced adhesiveness (see Fig.~4C, main text) indicating that the reinforcement of the effect of polarization on the migration dynamics through to the increased effective spring constants dominates. Directly varying the biological parameters that affect the polarization dynamics in a controlled way is experimentally difficult. The fit of our model to the observed migration biases of HT-1080 cells (Fig.~4B, main text) indicates however a weaker polarization compared to MDA-MB-231 cells. In our model, this leads to a stronger impact of the asymmetries in contractility on the overall migration dynamics resulting the migration being more biased towards the wider channel, where cells can form stronger protrusions (see fit in Fig.~4B, main text).

Based on these observations, we propose that the decision making of cells when encountering confinements of different widths, can be qualitatively understood in a reduced two dimensional phase space, where one dimension characterizes the overall effective contractility and the second dimension characterizes the degree of polarization. To estimate the position of the different experiments shown in Fig.~4B and C in the main text in this reduced parameter space, we choose the value of $k_{r/\ell}/\zeta_a$ to determine the position in the effective contractility dimension and the value of $n_cc_0$ to be representative of the position along the polarization dimension, since it contributes to all parameters determining the polarization dynamics (see Section 'Differences between cell lines' for a more detailed discussion of the polarization dimension). A sketch of this reduced phase space together with the estimated positions of the different island chain experiments is shown in Fig.~\ref{fig:figS2}B.

\subsection{Difference between cell lines}
Since the parameters used to describe MDA-MB-231 and HT-1080 cells on island chains were constrained from the same type of data, they lend themselves best to discuss differences between cell lines. We observe two trends when comparing parameter values: The rescaled spring constants are slightly lower for the HT-1080 cells than for the MDA-MB-231 cells and parameters associated with the polarization dynamics are strongly reduced in HT-1080 cells compared to MDA-MB-231 cells. The differences in rescaled spring constants could originate from different adhesive properties of the cell line or differences in the actin and myosin expression levels. The more striking difference between the two cell lines seems however to be the difference in polarization, where all parameters associated with the polarization dynamics are reduced by about a factor of two. As discussed above, there are three obvious biological parameters that have an impact on the polarization dynamics:  $\tilde{D}$, $c_0$ and $n_c$, where $c_0$ only occurs in combination with $n_c$ (see Eq.~(8), main text). One possible explanation for the observed difference in all polarization associated parameters could thus be an overall reduced concentration of polarity cue in HT-1080 cells compared to MDA-MB-231 cells. Interestingly, also $D - 8n_c c_0 \ell_a r_1L_c^{-1}$ is reduced by about a factor of two even though $D = 4(1-n_c)\tilde{D}L_c^{-2}$ does not directly depend on $c_0$. This could indicate a coupling between the overall concentration of the polarity cue $c_0$ and the bound fraction of the polarity cue $n_c$. A possible mechanism behind that could be that with a lower absolute number of polarity cue, there are more unoccupied binding sites available on the actin filaments, leading to an increase of $n_c$ and thus a decrease of $D$ with decreasing $c_0$. A more detailed characterization of the molecular differences between MDA-MB-231 cells and HT-1080 cells would however be needed to understand the observed differences in parameter values describing these two cell lines.

\newpage
\section{Approximate model with a single protrusion coordinate}
Starting from Eqs.~(4)~and~(5) in the main text, we want to derive from basic biophysical principles a data-driven model we found previously for mesenchymal cell migration inferred from experiments on dumbbell-shaped micro-patterns~\cite{bruckner_geometry_2022}. This model, however, only considers a single protrusion coordinate 
\begin{equation}\label{eq:protrusion_coordinate}
x_p = \frac{\Delta x_\ell x_\ell + \Delta x_r x_r}{\Delta x_\ell + \Delta x_r}, 
\end{equation} 
where $\Delta x_{\ell/r}$ denotes the growth of the respective protrusion between two consecutive experimental observation times spaced by $\Delta t$. While $\Delta x_{\ell/r}$ exhibit an intricate position and time dependence, we can show that we obtain the model found in~\cite{bruckner_geometry_2022} as an approximation of the more detailed model derived in the main text. 

\begin{figure*}[hbt]
\includegraphics{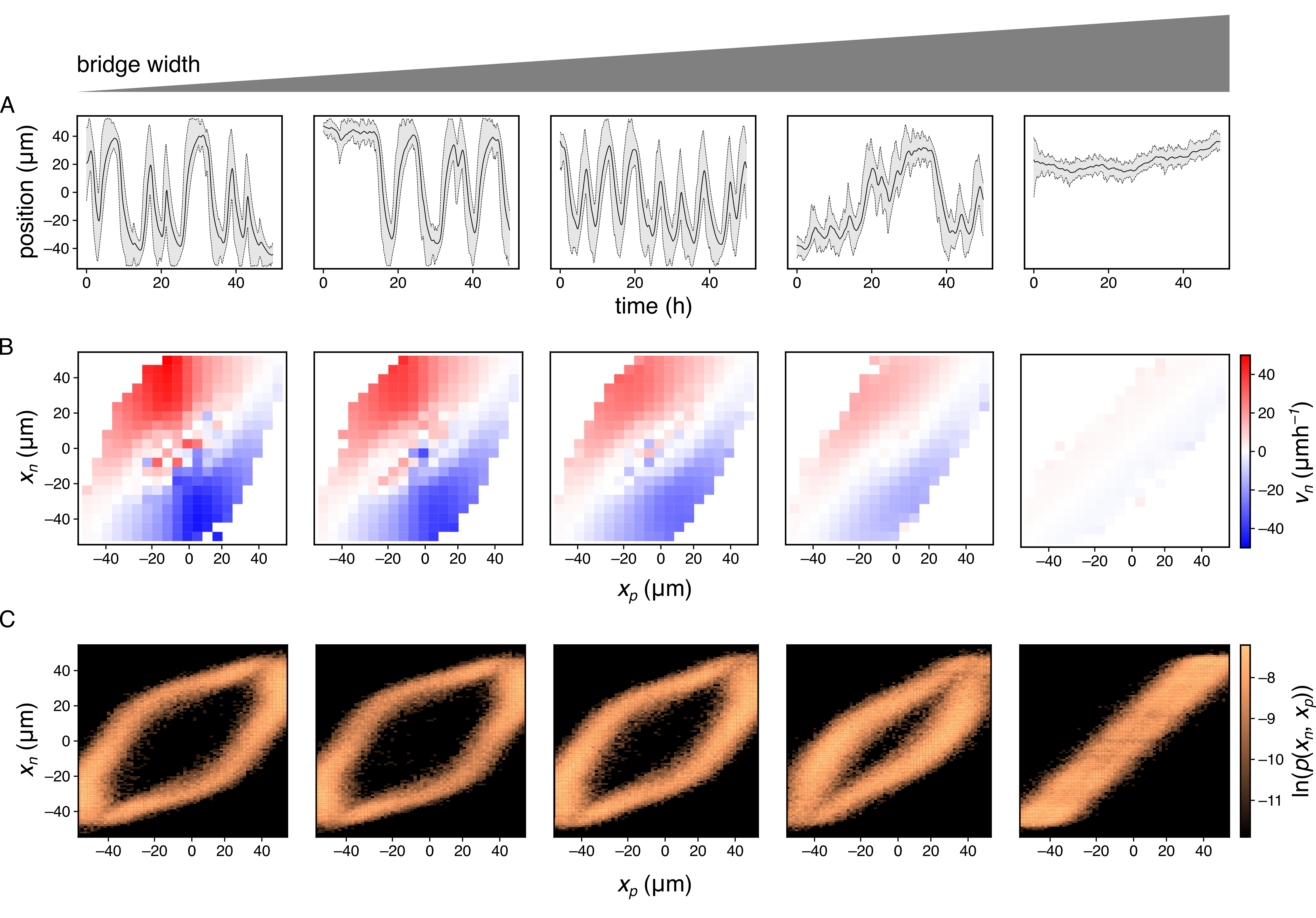}
\caption{\label{fig:figS3} \textbf{Dynamics of the two protrusion model for varying bridge widths (4\textmu m, 7\textmu m, 12\textmu m, 22\textmu m and 35\textmu m).} \textbf{A.} Example trajectories. \textbf{B.} Nuclear velocity maps. \textbf{C.} Joint probability densities.}
\end{figure*}

Assuming that the dynamics of $\Delta x_{\ell/r}$ are slower than the dynamics of $x_{\ell/r}$, we can approximate the dynamics of the protrusion coordinate as
\begin{equation}\label{eq:xp_dynamics}
\dot{x}_p = -\frac{k}{\zeta_a}(x_p - x_n) + \frac{\Delta x_r}{\Delta x_\mathrm{tot}}\ell_aS_r r_p(x_r) - \frac{\Delta x_\ell}{\Delta x_\mathrm{tot}}\ell_aS_\ell r_p(x_\ell),
\end{equation}
where $k = k_\ell = k_r$ and $\Delta x_\mathrm{tot} = \Delta x_\ell + \Delta x_r$. Eq.~\eqref{eq:xp_dynamics} contains a simple elastic coupling to the nucleus and a complex dependence on the projected polymerization rate in both protrusions. This second part is phenomenologically captured in~\cite{bruckner_geometry_2022} by introducing a confinement potential that can be interpreted as the outer boundaries of the micro-pattern prohibiting further actin polymerization and a polarization force $P(t)$ that essentially follows Eq.~(8) from the main text. In \cite{bruckner_geometry_2022}, the values of $\alpha$ in the center of the channel were determined by fitting the experimentally observed protrusion dynamics for each bridge width, while here we  obtain an analytical expression for the bridge width dependence of $\alpha(x_p)$ in Eq.~(9) in the main text. The found expression is consistent with the values obtained from fitting in \cite{bruckner_geometry_2022} (see Fig.~3B, main text). The main difference between Eq.~(8) in the main text and~\cite{bruckner_geometry_2022} is the inclusion of a bias term. Both models, however, lead to a qualitatively similar transition behavior (see Fig.~\ref{fig:figS3}).

Next, we consider the nuclear dynamics given by Eq.~(5) in the main text. Note that we can rewrite the prefactors as
\begin{equation}\label{eq:nuclear_pref}
\frac{k_c(x_{\ell/r})}{\zeta_n(w_n)} = \frac{k_c(x_{\ell/r})}{\zeta_n(w_{\rm cell})} \gamma(x_n)^{-1},
\end{equation}
where $\zeta_n(w_n)$ is given by Eq.~\eqref{eq:nuclear_friction} and $w_{\rm cell}$ denotes the unconfined width of the cell. The first term in Eq.~\eqref{eq:nuclear_pref} can be interpreted as the rescaled spring constant used in~\cite{bruckner_geometry_2022}, while the second term is equivalent to the phenomenological friction coefficient in~\cite{bruckner_geometry_2022}.

Note that the rescaled spring constant in Eq.~\eqref{eq:nuclear_pref} depends on the position of the protrusion, while it is assumed to be constant in~\cite{bruckner_geometry_2022}. However, since the protrusion coordinate strongly fluctuates, it is reasonable to approximate $k_c(x_{\ell/r}) \approx \langle k_c(x_{\ell/r}) \rangle$, which can be assumed to be roughly constant. We can the rewrite Eq.~(5) of the main text as
\begin{equation}
\dot{x}_n \approx \gamma_n(x_n)^{-1}k_n(x_\ell+x_r-2x_n),
\end{equation}
with
\begin{equation}
k_n = \frac{\langle k_c(x_{\ell/r}) \rangle}{\zeta_n(w_{\rm cell})}.
\end{equation}

In the limiting cases of a strongly polarized cell towards the right $\Delta x_{\ell} + \Delta x_r \approx \Delta x_r$. In that case, $x_{\ell} \approx x_n$ and $x_r \approx x_p$, such that $x_\ell+x_r-2x_n \approx x_p - x_n$. Similarly, in the case of a completely unpolarized cell, $\Delta x_r = \Delta x_\ell$ and thus $x_\ell + x_r = 2 x_p$. At the same time, $x_p \approx x_n$, such that again $x_\ell+x_r-2x_n \approx x_p - x_n$. Hence, in those two limiting cases we recover the nuclear dynamics found in~\cite{bruckner_geometry_2022}.

\newpage
\section{Experimental methods}
\subsection*{Micropatterning and sample preparation}
To passivate the surface of the ibiTreat \textmu-dish (ibidi), a small drop of 0.01\% (w/v) PLL (Thermofisher) was applied onto the surface of the dish. After incubating the dish at room temperature for a duration of 30min, the PLL-coated dish was rinsed using HEPES buffer (Roche, pH=8.3) to ensure thorough cleansing and preparation. A solution of mPEG-SVA (LaysanBio) with a concentration of 100mg/ml, diluted in 0.1M Hepes, was uniformly distributed onto the PLL-coated dish. The dish was then subjected to an incubation period for at least 1h at room temperature, allowing for appropriate bonding. Following this incubation, the dish was rinsed thoroughly with milliQ water to remove any residual substances. The passivated dish was then photopatterned using the PRIMO module (Alvéole), which was integrated into an automated inverted microscope (Nikon Eclipse Ti). This photopatterning process involved the utilization of the photoactivable reagent PLPP (Alvéole). To ensure uniform distribution of the PLPP gel, it was appropriately diluted in 99\% ethanol before being applied onto the passivated surface of the dish. The placement of the dumbbell-chain pattern onto the dish was achieved using the Leonardo software (Alvéole), and subsequently, it was exposed to UV-light with a dose of 15mJ/mm². After the illumination process, the dish was thoroughly washed with milliQ water and rehydrated with PBS for 5min. Following rehydration, the dish was incubated with either 10\textmu g/ml or 20\textmu g/ml of labeled Fibronectin-Alexa647 (provided by Y-proteins, Thermofisher) for 15 minutes at room temperature. Samples are stored in PBS at room temperature until cell seeding.

\subsection*{Cell culture}
MDA-MB-231 human breast carcinoma epithelial cells, co-expressing fluorescently labeled histones (mcherry-H2B), are cultured in common growth medium L-15 (Sigma) supplemented with 10\% foetal bovine serum (Sigma). The cells are cultivated at a temperature of 37°C up to 80-90\% confluence.
HT1080 fibrosarcoma cells are cultured in MEM (Sigma) supplemented with 10\% foetal bovine serum (Sigma). The cells are cultured at 37°C in a 5\% CO$_{2}$ environment with humidity maintained until reaching 80-90 \% confluence.\\
Following this, the cells were washed and trypsinized for 4min. For experimental purposes, the cell solution is centrifuged at 1,000 r.c.f. for 3min. Subsequently, the cells were re-suspended in L-15. Approximately 8,000 cells were seeded per $\mu$-dish (ibidi), allowing them to adhere for a minimum duration of 3h. To achieve nuclear staining of HT1080 cells, 15 nM Hoechst 33342 is added. Throughout the experiments, cells are maintained within a humidified environment at 37°C.

\subsection*{Microscopy and cell tracking}
Measurements were performed in time-lapse mode for up to 48h on a Nikon-Eclipse TI-E inverted microscope. To provide standard incubation conditions throughout the measurements, the microscope is equipped with gas incubation and a heating system (Okolab). Bright-field and fluorescence images of the fibronectin-coated pattern and the co-expressed labeled histones were acquired every 10min. A bandpass filter is applied to the images of the nuclei to enhance their quality. Following this, the images were binarized, and the positions of the nuclei's center-of-mass were determined using the Analyze Particles plugin in ImageJ.

\bibliography{references}% Produces the bibliography via BibTeX.